\pdfoutput=1
\documentclass[11pt]{article}
\usepackage{graphicx}
\usepackage{hyperref}
\usepackage{booktabs}
\usepackage{array}
\usepackage{amsmath}
\usepackage{amssymb}
\usepackage{anysize}
\usepackage[numbers,comma,square,sort&compress]{natbib}
\usepackage[font=small, labelfont=bf]{caption}

\newcolumntype{C}{>{\centering\arraybackslash}m{3.2cm}}

\hypersetup{%
    pdftitle = {Modeling crosstalk in silicon photomultipliers},
    pdfsubject = {preprint},
    pdfkeywords = {SiPM, Crosstalk, Afterpulsing, Excess noise factor},
    pdfauthor = {L. Gallego, J. Rosado, F. Blanco and F. Arqueros},
    pdfstartview = {FitH},
    }

\marginsize{3cm}{3cm}{2cm}{2cm}

\begin{document}
\begin{titlepage}
\begin{center}
\vspace*{3.0cm}\Huge {\bf Modeling crosstalk in silicon photomultipliers}
\par
\vspace*{2.0cm} \normalsize {\bf {L.~Gallego, J.~Rosado,  F.~Blanco and F.~Arqueros}}
\par
\vspace*{0.5cm} \small \emph{Departamento de F\'{i}sica At\'{o}mica, Molecular y Nuclear, Facultad de Ciencias
F\'{i}sicas, Universidad Complutense de Madrid, E-28040 Madrid, Spain}
\end{center}
\vspace*{2.0cm}

\begin{abstract}
Optical crosstalk seriously limits the photon-counting resolution of silicon photomultipliers. In this work, realistic
analytical models to describe the crosstalk effects on the response of these photodetectors are presented and compared
with experimental data. The proposed models are based on the hypothesis that each pixel of the array has a finite
number of available neighboring pixels to excite via crosstalk. Dead-time effects and geometrical aspects of the
propagation of crosstalk between neighbors are taken into account in the models for different neighborhood
configurations. Simple expressions to account for crosstalk effects on the pulse-height spectrum as well as to
evaluate the excess noise factor due to crosstalk are also given. Dedicated measurements were carried out under both
dark-count conditions and pulsed illumination. Moreover, the influence of afterpulsing on the measured pulse-height
spectrum was studied, and a measurement of the recovery time of pixels was reported. High-resolution pulse-height spectra
were obtained by means of a detailed waveform analysis, and the results have been used to validate our
crosstalk models.

\end{abstract}
\end{titlepage}

\section{Introduction}
\label{intro}

Silicon photomultipliers (SiPMs) belong to the sort of photodetectors with single-photon detection capability
\cite{Renker}. These devices consist in a monolithic array of multiple single avalanche photodiodes (APDs) operated in
Geiger mode. Each component or pixel produces a pulse of constant amplitude regardless of the number of impinging
photons. All pixels are connected to the same output channel providing a total output signal equal to the sum of those
from the individual pixels, which enables a large dynamic range and photon counting. SiPMs offer desirable properties
like high gain and sensitivity, and very good time and photon-counting resolutions, which have increased interest
in the use of these photodetectors in numerous applications displacing more traditional ones. Unfortunately, their
photon-counting ability is seriously limited by optical crosstalk, which affects the linearity of the detector, in
particular, causing a significant excess noise. The very relevant influence of crosstalk on the performance of SiPMs
has been contemplated in most characterization studies of these devices. However, a more accurate description of this
effect is still necessary.

When a pixel is fired by either an incoming photon or by a thermally generated electron-hole pair, hot carriers in the
avalanche breakdown induce emission of IR photons \cite{Bude,Lacaita,Villa} that in turn may trigger further avalanches
in nearby pixels. This stochastic process, called optical crosstalk, is characterized by being nearly instantaneous,
and its probability is proportional to the SiPM gain. Operating at low bias voltage would diminish significatively
crosstalk effects, but at the expense of degrading the photon-detection efficiency. The incorporation of isolation
trenches around each pixel, as proposed by \cite{Mathewson}, successfully reduces optical crosstalk
\cite{Kindt,Buzhan,MacNally}. This technique has become very usual in the fabrication of this kind of photodetectors.
However, it was first shown in \cite{Rech} that a very significant contribution to crosstalk can come from light
reflected on the bottom surface of the Si bulk, and thus, trenches would not be able to completely prevent it.

The main effect of crosstalk is to introduce a multiplication noise, although it does not affect the pulse-height
resolution. So, at conditions where only one pixel is expected to be excited simultaneously (e.g., dark counts),
crosstalk results in output pulses with amplitudes twice or several times the amplitude of a single triggered pixel.
The crosstalk probability $\varepsilon$ is usually defined as the rate of dark counts with crosstalk (two or more fired
pixels) divided by the total dark-count rate.

Several studies based on Monte Carlo simulations \cite{SanchezMajos,Vacheret,Dovrat} have shown that the response of
various SiPM devices from Photonique and Hamamatsu can be properly described by assuming that crosstalk only takes
place between adjacent pixels. On the other hand, complete simulations \cite{Rech,Otte} of some non-commercial SiPM
devices, including light propagation through the silicon bulk and reflections on the bottom surface, have demonstrated
that crosstalk can also be induced in distant pixels, which is supported by experimental data for these SiPMs.

An analytical formulation is desirable for a better understanding of crosstalk effects. Most of the available
statistical models of crosstalk \cite{Eraerds,Afek,Vinogradov1,Ramilli,Dovrat2,Kalashnikov,vanDam,Putignano} are based
on simple assumptions involving restrictions on the number of crosstalk events per initially fired pixel or on
crosstalk cascading (i.e., crosstalk events generated by other crosstalk-induced avalanches). S.~Vinogradov has
developed very recently an analytical model overcoming these limitations, where crosstalk is considered as a branching
Poisson process \cite{Vinogradov2}. However, no analytical model has so far dealt with the geometrical arrangement of
the neighboring pixels available to be excited via crosstalk and the consequent local saturation effects as pixels
become inactive after being excited. In the present work, realistic analytical models of crosstalk that take into
account these geometrical considerations are reported and validated by comparison with experimental data.

This paper is organized as follows. In section~\ref{sec:models}, the proposed crosstalk models are explained, and
analytical expressions are provided for the probability distribution of the number of crosstalk events at both
dark-count and pulsed-illumination conditions. The experimental setup and the analysis method are described in
section~\ref{sec:experiment}. Results are presented in section~\ref{sec:results}. Conclusions are drawn in
section~\ref{sec:conclusions}.

\section{Analytical models of crosstalk}
\label{sec:models}

Next, fundamentals of the statistics of crosstalk events (including cascading effects) for a single initially fired
pixel are described. Then, realistic crosstalk models that take into account the above-mentioned geometry-dependent
saturation effects are detailed in section~\ref{ssec:saturation}. Finally, effects of crosstalk on the photon
statistics under pulsed-illumination conditions are discussed in section~\ref{ssec:random_primaries}.

\subsection{Fundamentals}
\label{ssec:fundamentals}

Several analytical models of crosstalk noise in SiPMs are available in the literature. Many of them
\cite{Eraerds,Afek,Vinogradov1,Ramilli,Dovrat2,Kalashnikov} assume that crosstalk obeys a Bernoulli distribution, that
is, a primary avalanche can either trigger a secondary avalanche in a neighboring pixel with probability $p$ or no
avalanche with probability $1-p$. Some of these models also include cascading processes where a secondary avalanche may
trigger a tertiary avalanche, which could in turn trigger a quaternary avalanche and so on, although the number of
crosstalk events that can be induced directly by each pixel is limited to one. However, a single avalanche is made up
of a large number of carriers (typical avalanche gain is $10^5 - 10^6$), each one being able to induce emission of
crosstalk photons with a probability of $\sim 3\cdot10^{-5}$ \cite{Lacaita,Otte,Mirzoyan}. Accordingly, some authors
\cite{vanDam,Putignano,Vinogradov2} have used a Poisson distribution instead to describe the number of crosstalk events
initiated directly by a single avalanche.

Even though many IR photons may be produced in a single avalanche breakdown, they are expected to be absorbed
preferably in the vicinity of the pixel in which the avalanche was triggered, and therefore, the number of neighboring
pixels able to be excited efficiently via crosstalk must likely be small. This has suggested to us to use a binomial
distribution to describe the number $s$ of successful crosstalk events in a finite sample of $n$ available neighbors of
the primary pixel, that is,

\begin{equation}
\label{binomial}
P(s)={{n}\choose{s}}p^s\,(1-p)^{n-s}\,,
\end{equation}

where the simple approximation of assuming the same probability $p$ of crosstalk for any individual neighbor is made.
This $p$ parameter is related to the total crosstalk probability $\varepsilon$ (one or more crosstalk events) through
$1-\varepsilon=(1-p)^n$. Note that (\ref{binomial}) reduces to the Bernoulli distribution for $n=1$ and tends to a
Poisson distribution with mean $\lambda=-\ln(1-\varepsilon)$ for $n\rightarrow\infty$ keeping constant $\varepsilon$.
The latter case would correspond to the limit situation where crosstalk photons originated from the primary pixel can
reach any other pixel of a very large array.

\begin{figure}[t]
\begin{center}
\includegraphics[width=.8\textwidth]{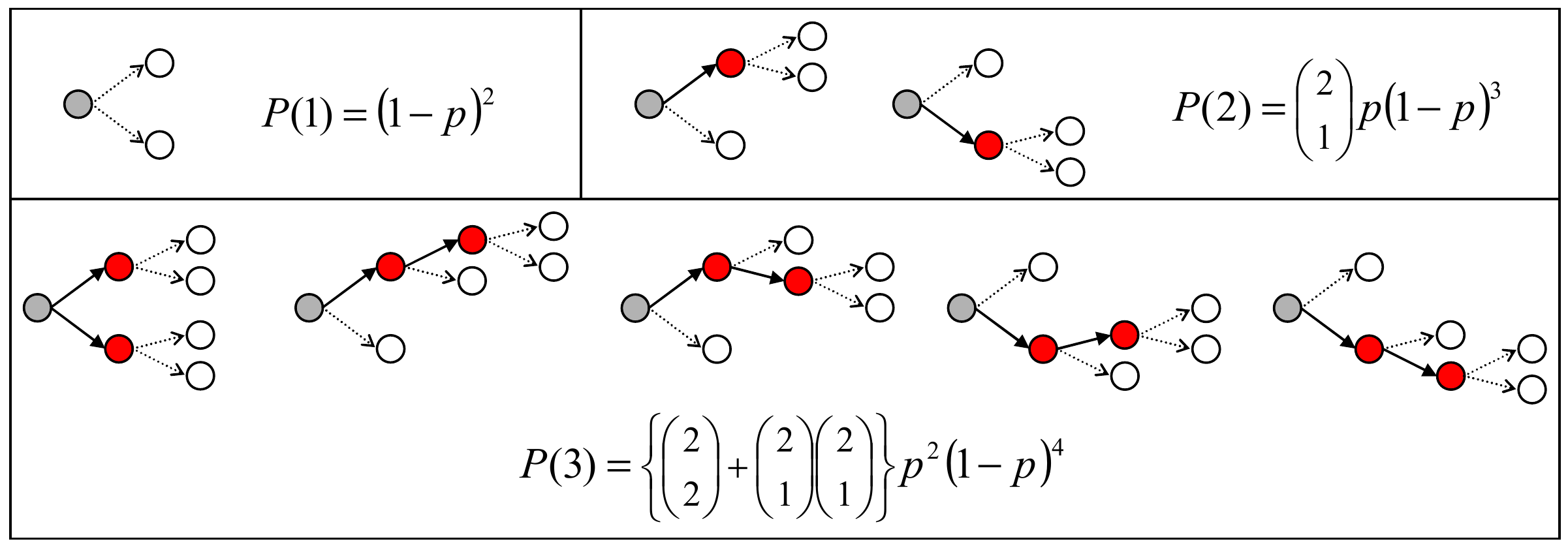}
\end{center}
\caption{Histories of crosstalk excitations for a model of 2 neighbors and total number of triggered pixels from 1 to 3.
The primary and the subsequent crosstalk events are represented by grey-filled and red-filled circles, respectively.
The cascade development is indicated by solid arrows, while a dotted arrow ending in an open circle means that
the cascade no longer continues because the neighbor is not triggered.
The corresponding probabilities are given in terms of combinatorial numbers.}
\label{fig:histories_2neighbors}
\end{figure}

To account for the total number of crosstalk events per initially fired pixel, crosstalk cascading should be included.
As a first approximation, we assume that any triggered pixel, either the primary pixel or a secondary or higher-order
one, can always induce crosstalk in $n$ more neighboring pixels. So, the probability of a cascade of crosstalk events
would be obtained by applying (\ref{binomial}) repeatedly for each pixel of the sequence. The probability to get a
certain total number $k$ of triggered pixels is proportional to the number of histories with $k-1$ crosstalk events,
considering all the possible combinations of neighbors that are excited or not in each step. This is illustrated in
figure~\ref{fig:histories_2neighbors}, where all the histories with up to two crosstalk events are identified for a
model of $n=2$ neighbors. For arbitrary $n$ and $k$ values, probabilities can be calculated as

\begin{equation}
\label{n_model}
P(k)=h_{n,k-1}\,p^{k-1}\,(1-p)^{k\,n-k+1}\,,
\end{equation}
where $h_{n,k-1}$ denotes the number of histories of crosstalk excitations adding up to $k-1$ for a model of $n$
neighbors. The count of histories is not trivial for a large number of triggered pixels. Nevertheless, the procedure
shown in figure~\ref{fig:histories_2neighbors} can be generalized by the recursive formula

\begin{equation}
\label{h_nc}
h_{n,c}=\sum_{s=1}^{c}{{n}\choose{s}}
\sum_{i_1=0}^{c-s}h_{n,i_1}\sum_{i_2=0}^{c-s-i_1}h_{n,i_2}\;\;\ldots\!\!\!\sum_{i_{s-1}=0}^{c-s-i_1-\ldots-i_{s-2}}\!\!\!h_{n,i_{s-1}}
\,h_{n,c-s-i_1-\ldots-i_{s-1}}\,,
\end{equation}
starting with $h_{n,0}=1$. This results in the following probabilities for $k\leq 5$:

\begin{align}
\label{n_model_k5}
&P(1)=(1-p)^n=1-\varepsilon \nonumber\\
&P(2)=n\,p(1-p)^{2n-1} \nonumber\\
&P(3)=\frac{1}{2}n(3n-1)p^2(1-p)^{3n-2} \nonumber\\
&P(4)=\frac{1}{3}n\left(8n^2-6n+1\right)p^3(1-p)^{4n-3} \nonumber\\
&P(5)=\frac{1}{4}n\left(\frac{125}{6}n^3-25n^2+\frac{55}{6}n-1\right)p^4(1-p)^{5n-4}\,.
\end{align}
Notice that the overall crosstalk probability $\varepsilon$ is unaffected by including or not cascading effects,
because it only relies on the original binomial probability that the primary pixel induces no crosstalk event.

The mean and variance of the distribution are conveniently approximated as power series of either $p$ or $\varepsilon$
up to second order\footnote{Both mean and variance grow rapidly, and even diverge, at increasing $p$ and $\varepsilon$
values. As a consequence, this approximation fails at large crosstalk probability ($\varepsilon>0.2$).}

\begin{equation}
\label{mean_n_model}
E=1+n\,p+(n\,p)^2+o\left(p^2\right)=1+\varepsilon+\frac{3n-1}{2n}\varepsilon^2+o\left(\varepsilon^2\right)
\end{equation}
\begin{equation}
\label{variance_n_model}
Var=n\,p+n(3n-1)p^2+o\left(p^2\right)=\varepsilon+\frac{7n-3}{2n}\varepsilon^2+o\left(\varepsilon^2\right)\,.
\end{equation}

\begin{figure}[t]
\begin{center}
\includegraphics[width=.4\textwidth]{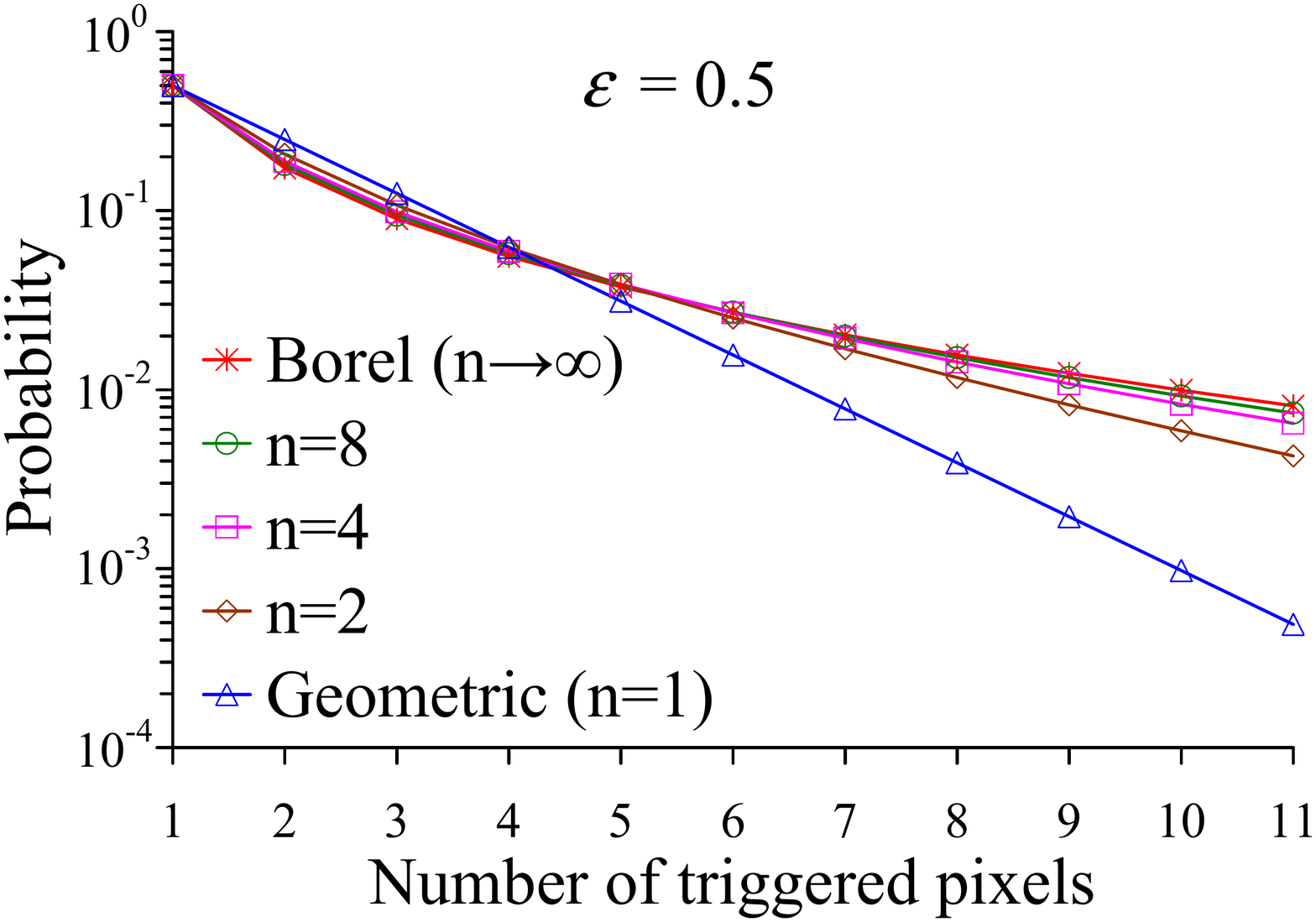}
\includegraphics[width=.4\textwidth]{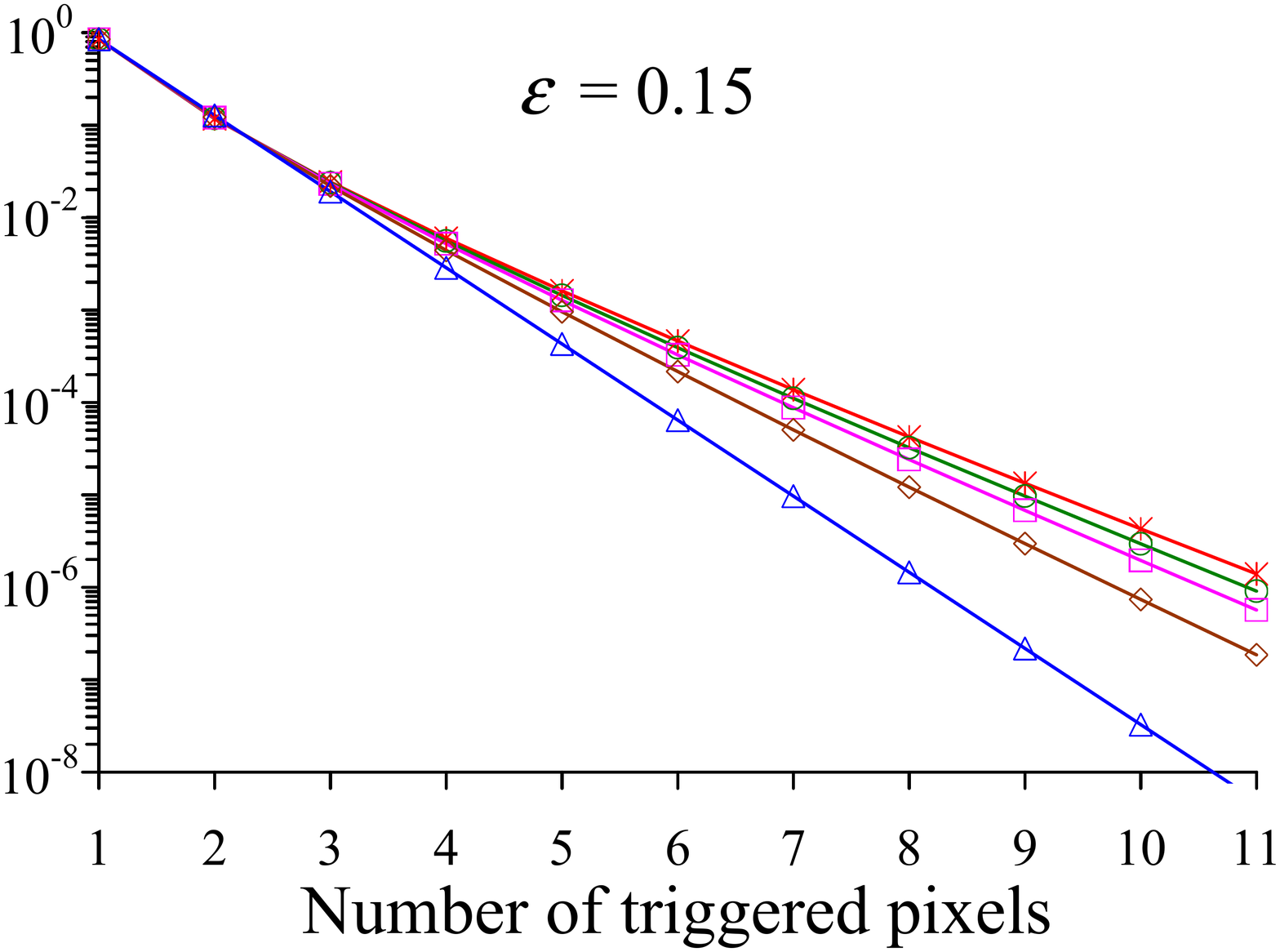}
\end{center}
\caption{Probability distributions of the total number of triggered pixels for a single initially fired pixel
resulting from equations (\protect\ref{n_model}) and (\protect\ref{h_nc}) for different numbers of neighbors $n$. The
total crosstalk probability (one or more crosstalk events) is set to $\varepsilon=0.5$ (left) and to $\varepsilon=0.15$
(right).}
\label{fig:n_models}
\end{figure}

In figure~\ref{fig:n_models}, results of equations (\ref{n_model}) and (\ref{h_nc}) are shown for two different
crosstalk probabilities $\varepsilon=0.5$ (left) and $\varepsilon=0.15$ (right) and for several $n$ values. The
probability distributions for $n=1$ and for $n\rightarrow\infty$ correspond respectively to the geometric and Borel
distributions, which have already been exploited by S.~Vinogradov et~al. to model crosstalk
\cite{Vinogradov1,Vinogradov2}. For intermediate $n$ values, the probability distribution is between these two limit
situations, although it tends rapidly to the Borel distribution as $n$ increases. In general, the larger the number of
neighbors is, the more likely multiple crosstalk excitations are.

Crosstalk can be considered as an extra multiplication process introducing noise. This is usually characterized by the
excess noise factor according to the standard definition

\begin{equation}
\label{ENF}
ENF=1+\frac{Var}{E^2}\,,
\end{equation}
where $E$ represents the mean multiplication factor (gain) of the signal due to crosstalk for a single initially fired
pixel. Using~(\ref{mean_n_model}-\ref{variance_n_model}), the excess noise factor for a crosstalk model of $n$
neighbors can be approximately expressed as

\begin{equation}
\label{ENF_n_model}
ENF=1+n\,p+n(n-1)p^2+o\left(p^2\right)=1+\varepsilon+\frac{3n-3}{2n}\varepsilon^2+o\left(\varepsilon^2\right)\,.
\end{equation}
This parameter increases as $n$ increases at constant $\varepsilon$, as expected, although it saturates at a value of
$1/[1+ln(1-\varepsilon)]$, which corresponds to the Borel distribution.

\subsection{Models including saturation effects}
\label{ssec:saturation}

The frame of a binomial crosstalk probability (\ref{binomial}) with $n$ neighbors per pixel is our starting point to
construct a realistic crosstalk model. Probability distributions resulting from equations (\ref{n_model}-\ref{h_nc})
account for cascading of crosstalk; on the other hand they do not include saturation effects due to the fact that a
pixel that has already been triggered is inactive during a short period of time.

\begin{figure}[t]
\begin{center}
\includegraphics[width=.2\textwidth]{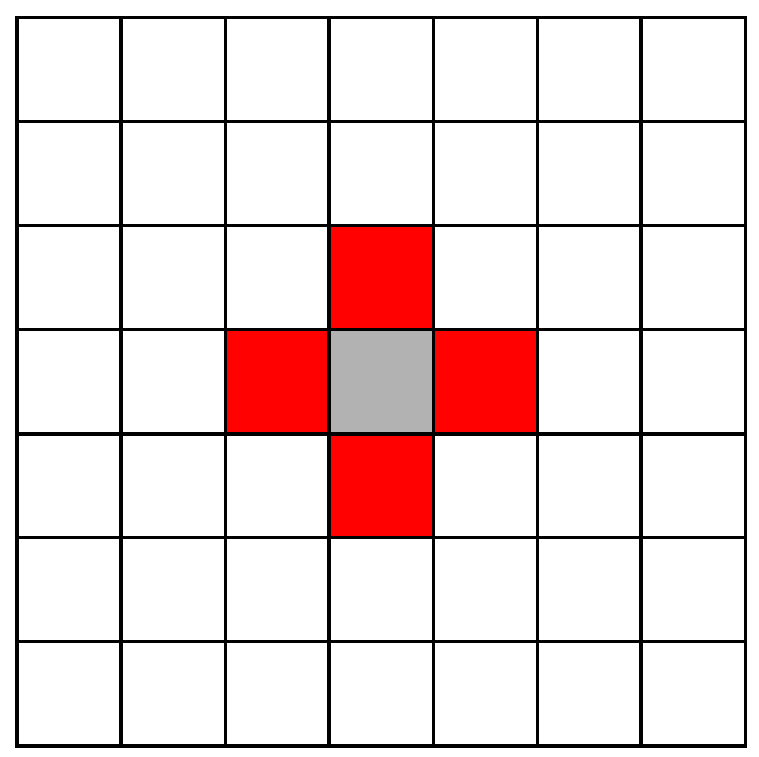}
\includegraphics[width=.2\textwidth]{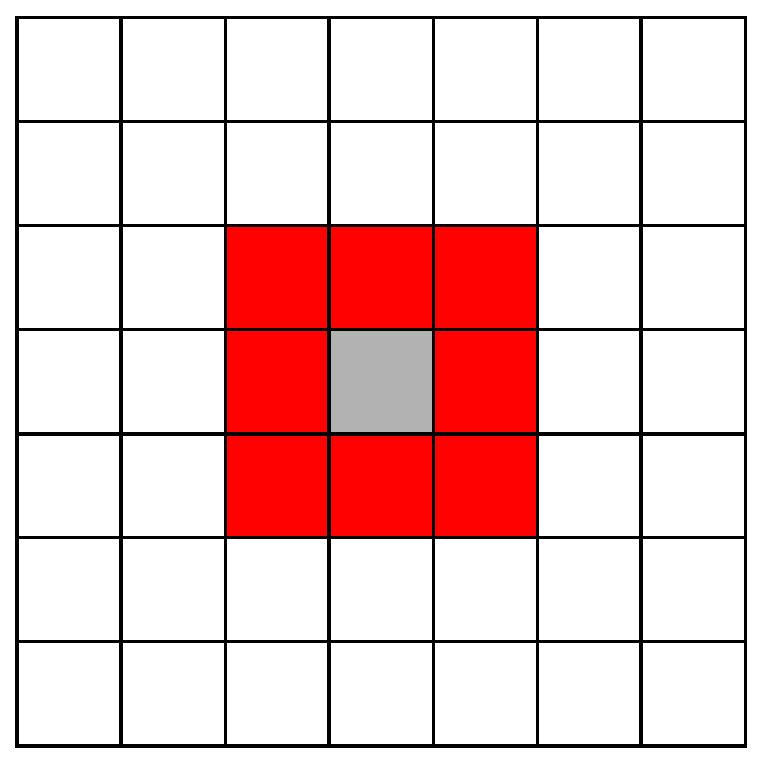}
\includegraphics[width=.2\textwidth]{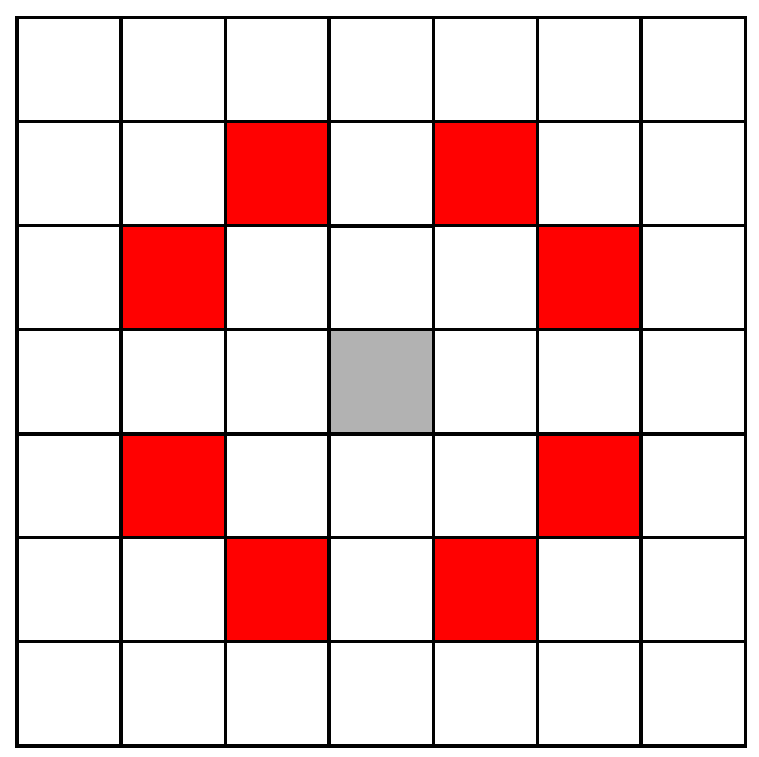}
\includegraphics[width=.2\textwidth]{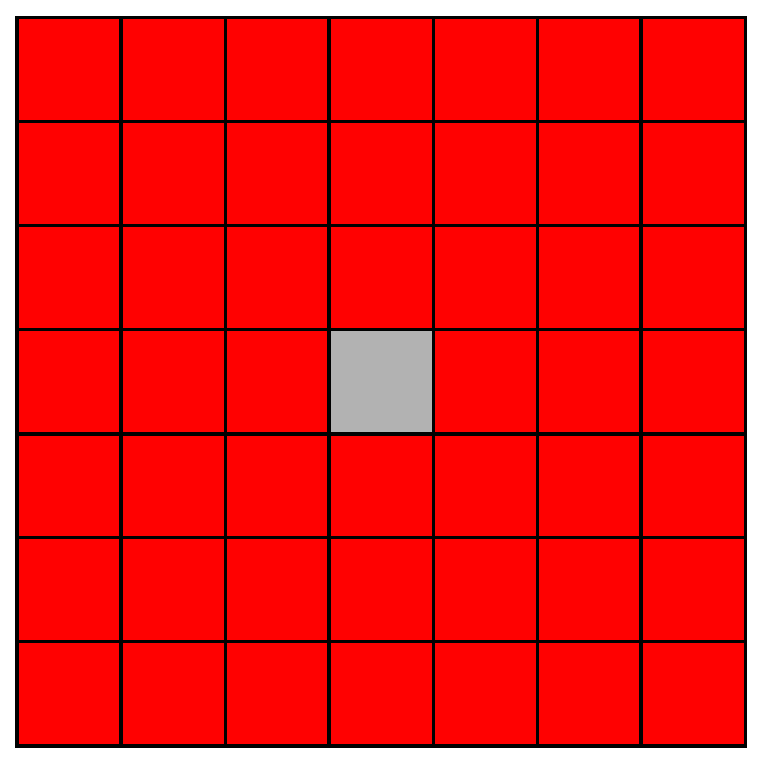}
\end{center}
\caption{Neighborhoods defined in the proposed geometrical crosstalk models including saturation effects. From left to right:
``4 nearest neighbors'', ``8 nearest neighbors'', ``8 L-connected neighbors'' and ``all neighbors''.}
\label{fig:geom_models}
\end{figure}

Different pixels can share neighbors in such a way that the last pixels in a history of excitations have less chances
to induce further crosstalk events. In fact, if pixel $i$ is a neighbor of pixel $j$, pixel $j$ is also a neighbor of
pixel $i$, but crosstalk between both pixels can only take place in one direction at a time: $i\rightarrow j$ or
$j\rightarrow i$. Thus, in a model of $n$ neighbors, only the primary pixel actually has the full number of neighbors
available to excite. This saturation effect will depend on how dense the neighborhood of a pixel is. To evaluate it in
an analytical way, we have considered four geometrical models of crosstalk defining different neighborhoods in a square
lattice of pixels: ``4 nearest neighbors'', ``8 nearest neighbors'', ``8 L-connected neighbors'' and ``all neighbors''
(figure~\ref{fig:geom_models}).

The 4-nearest-neighbors hypothesis has already been used successfully in Monte Carlo simulations of the response of
SiPMs including the crosstalk effect \cite{SanchezMajos,Vacheret,Dovrat}. However, as pointed out above, some works
have shown that crosstalk is also possible, or even more probable, between non-contiguous pixels \cite{Rech,Otte},
which has motivated to evaluate the other models. Note that the range of IR photons reflected on the surface of the
detector is expected to be strongly dependent on the particular chip configuration (e.g., pixel separation, width of
the Si bulk and depth of isolation trenches), and therefore, crosstalk effects in different SiPMs can be better
described by different geometrical models.

Unfortunately, simple expressions of the probability distribution of arbitrary number $k$ of triggered pixels cannot be
derived for such models. Nevertheless, only a few crosstalk excitations per initially fired pixel are likely to occur
at moderate crosstalk probability, and it may suffice, in practice, to determine the first four or five probabilities
of the distribution. The probability of $k=1$ (i.e., no crosstalk event) is trivially given by
$P(1)=(1-p)^n=1-\varepsilon$. To calculate the probabilities of $k\geq2$, one can proceed as follows:

\begin{enumerate}
\item All the different histories with $k-1$ crosstalk events are identified for the actual model taking into
    account the geometrical arrangement of neighbors and pixel inactivation.
\item The probability of each history is calculated as $p^c\,(1-p)^f$, where $c=k-1$ is the number of crosstalk
    events and $f$ is the overall number of crosstalk fails considering the active neighbors left for each
    triggered pixel following excitation order.
\item The final probability $P(k)$ is computed as the sum of probabilities of all the contributing histories.
\end{enumerate}

\begin{figure}[t]
\begin{center}
\includegraphics[width=.2\textwidth]{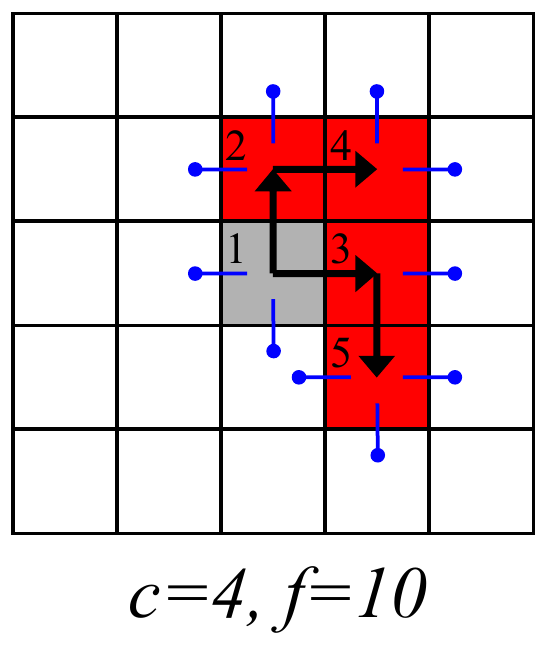}
\includegraphics[width=.2\textwidth]{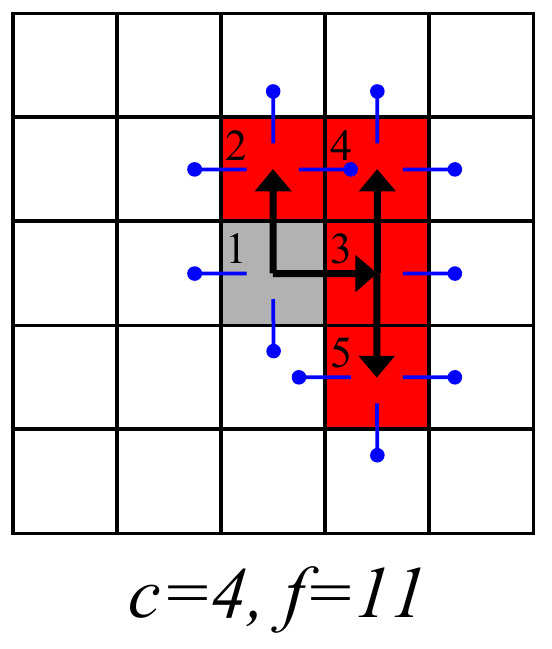}
\includegraphics[width=.2\textwidth]{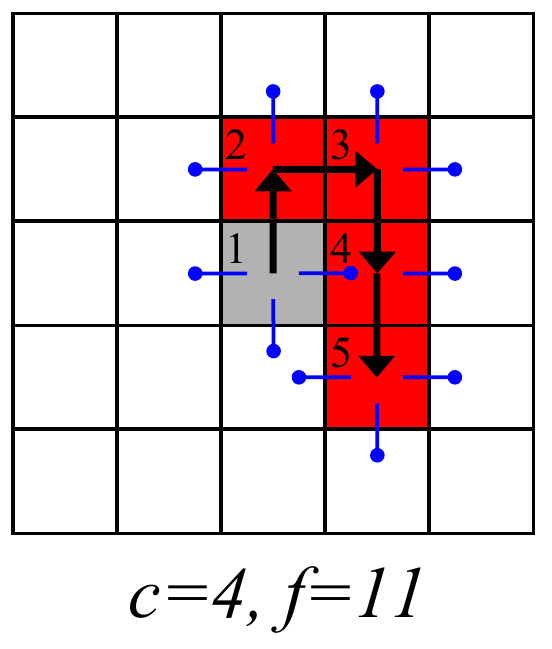}
\includegraphics[width=.2\textwidth]{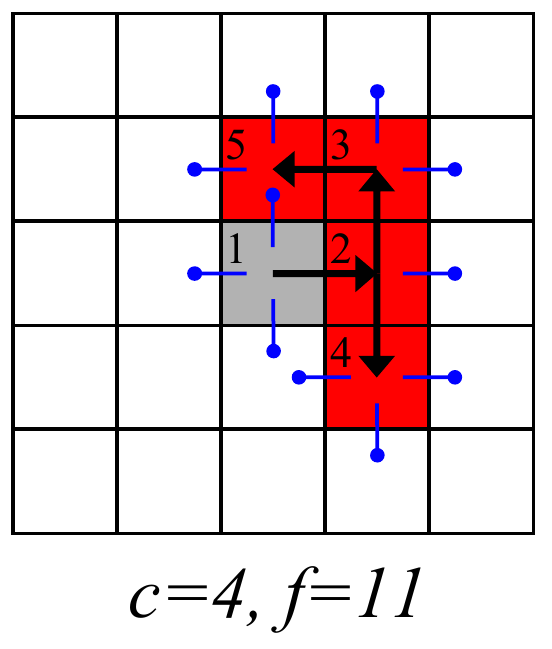}
\end{center}
\caption{Several crosstalk histories ($k=5$) giving rise to the same pattern of excitations in the 4-nearest-neighbors model. Both the number
$c$ of crosstalk excitations (arrows) and the number $f$ of crosstalk fails (short lines ending with dots) are indicated.
Excitation order (see labels on the corner of triggered pixels) is followed to determine the active neighbors left for each pixel.
To avoid ambiguities, crosstalk excitations induced by the same parent are assumed to be ordered clockwise starting from the top pixel.}
\label{fig:histories}
\end{figure}

For instance, assuming the 4-nearest-neighbors model, there are four symmetrical histories with $k=2$, where the
primary pixel and one of its neighbors are triggered, but neither the remaining three neighbors of the primary pixel
nor the other three neighbors left for the secondary one are excited by crosstalk (i.e., $f=6$). Therefore,
$P(2)=4p\,(1-p)^6$ is obtained. For $k=3$, 18 histories can still easily be identified, all of them with $f=8$,
resulting in $P(3)=18p^2\,(1-p)^8$. Calculations become involved for larger $k$ values, since the number of histories
increases rapidly, and they can also have different number $f$ of crosstalk fails due to the fact that pixels share
neighbors. To illustrate this, a few histories with $k=5$ are shown in figure~\ref{fig:histories}.

A computer algorithm was made to sequence the histories with $k\leq 5$ for the 4-nearest-neighbors, 8-nearest-neighbors
and 8-L-connected models. An endless array of pixels was assumed, that is, border effects were ignored. Collecting
histories with same $k$ value, analytical expressions of the corresponding $P(k)$ probabilities were obtained for these
models. For the all-neighbors model, probabilities of $k\leq 5$ were also determined as a function of the $p$ parameter
and the number $N$ of pixels in the array using combinatorial mathematics. Results are listed in
table~\ref{tab:probabilities}.

\begin{table}[t]
\caption{Probabilities of $k\leq 5$ triggered pixels for the different proposed models. The $q$ parameter is defined as
$1-p$ and $N$ is the total number of pixels in the array (last column).}\label{tab:probabilities} \footnotesize
\begin{tabular}{c C C C C}
\addlinespace
\toprule
$k$ & 4 nearest neighbors               & 8 nearest neighbors              & 8 L-connected neighbors           & All neighbors                         \\
\midrule
1   & $q^4 \left(=1-\varepsilon\right)$ & $q^8\left(=1-\varepsilon\right)$ & $q^8 \left(=1-\varepsilon\right)$ & $q^{N-1} \left(=1-\varepsilon\right)$ \\
\midrule
2   & $4p\,q^6$                         & $8p\,q^{14}$                     & $8p\,q^{14}$                      & ${{N-1}\choose{1}}p\,q^{2(N-2)}$      \\
\midrule
3   & $18p^2\,q^8$                      & $12p^2\,q^{18}\left[1+2q+4q^2\right]$
                                                                           & $84p^2\,q^{20}$                   & ${{N-1}\choose{2}}p^2\,q^{3(N-3)}
                                                                                                                 \left[1+2q\right]$                    \\
\midrule
4   & $4p^3\,q^8 \left[1+3q+18q^2\right]$
                                        & $4p^3\,q^{20} \left[1+3q\right.$ $+14q^2+30q^3+61q^4$ $\left.+59q^5+72q^6\right]$
                                                                           & $24p^3\,q^{24} \left[1+3q+38q^2\right]$
                                                                                                               & ${{N-1}\choose{3}}p^3\,q^{4(N-4)}$
                                                                                                                 $\left[1+3q+6q^2+6q^3\right]$         \\
\midrule
5   & $5p^4\,q^{10} \left[8+24q+55q^2\right]$
                                        & $5p^4\,q^{24} \left[9+36q+98q^2\right.$ $+188q^3+310q^4+372q^5$ $\left.+520q^6+396q^7+341q^8\right]$
                                                                          & $4p^4\,q^{30}$ $\left[180+540q+2521q^2\right]$
                                                                                                               & ${{N-1}\choose{4}}p^4\,q^{5(N-5)}$
                                                                                                                 $\left[1+4q+10q^2+20q^3\right.$
                                                                                                                 $\left.+30q^4+36q^5+24q^6\right]$     \\
\bottomrule
\end{tabular}
\end{table}

\begin{figure}[t]
\begin{center}
\includegraphics[width=.4\textwidth]{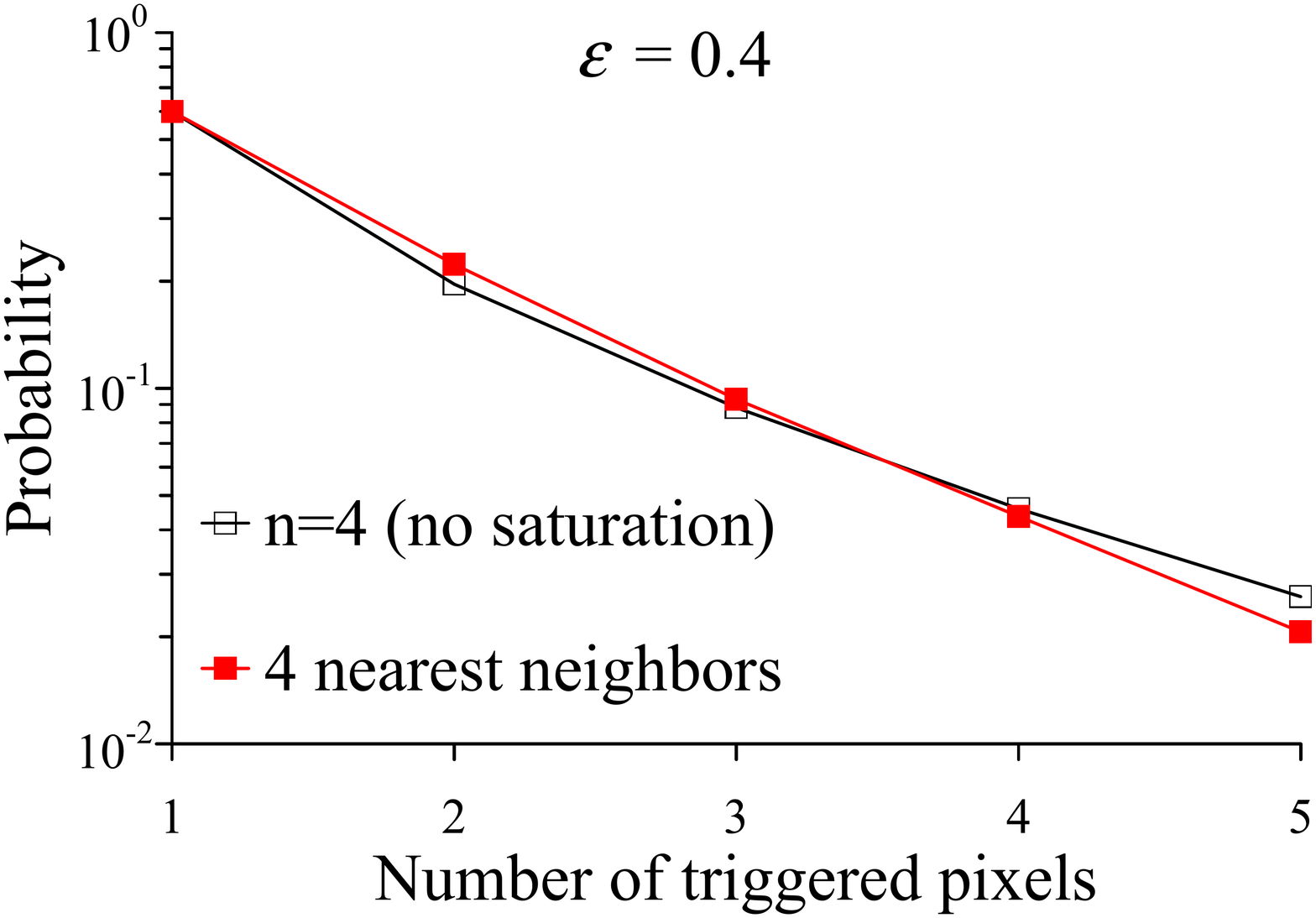}
\includegraphics[width=.4\textwidth]{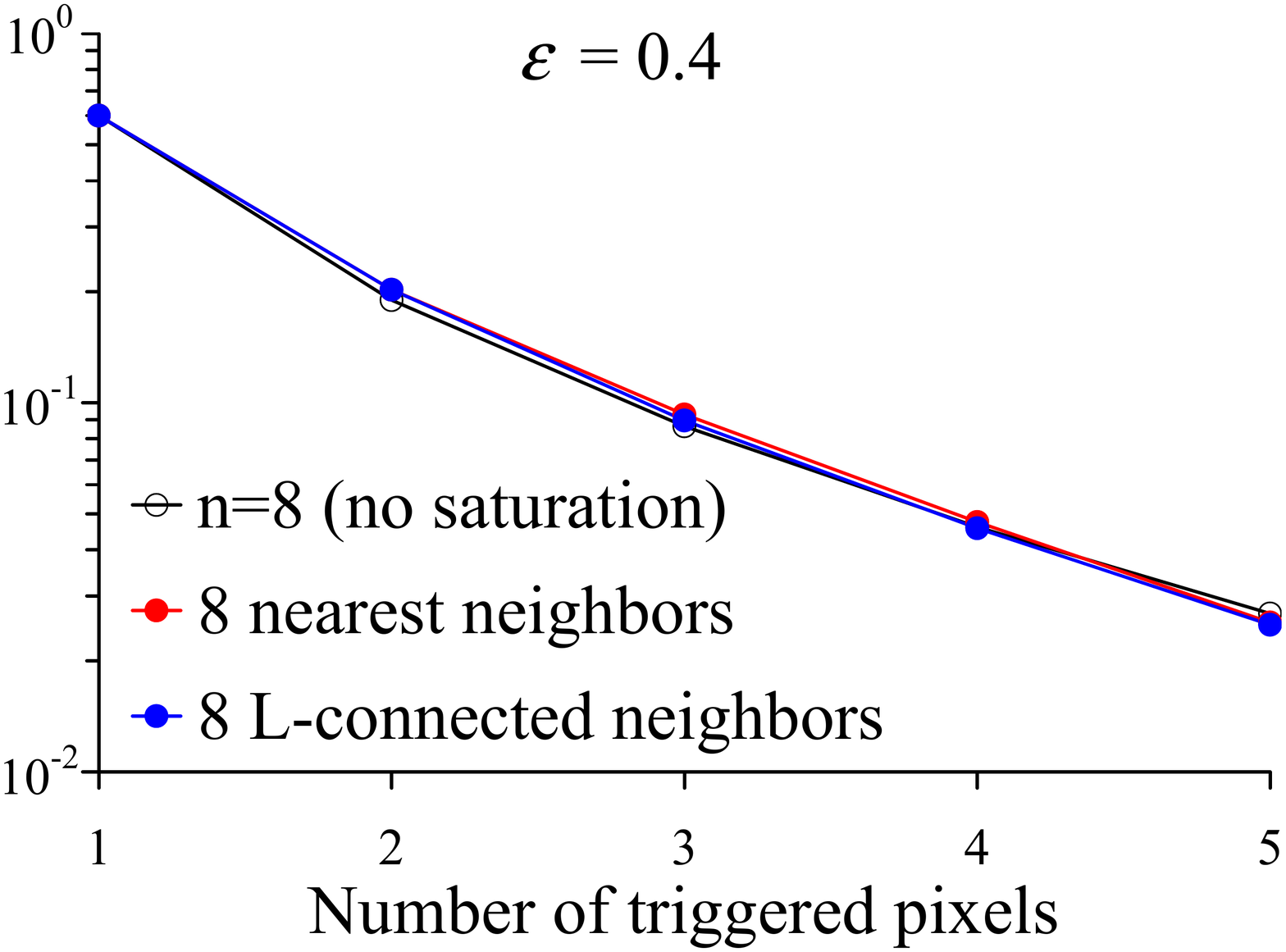}
\end{center}
\caption{Probability distributions for the 4-nearest-neighbors model (left) and for both the 8-nearest-neighbors and
8-L-connected-neighbors models (right) with $\varepsilon=0.4$. Results are compared with the distributions of
equations (\protect\ref{n_model}-\protect\ref{h_nc}) for $n=4$ and $n=8$, respectively, to illustrate the saturation effects.}
\label{fig:saturation}
\end{figure}

Figure~\ref{fig:saturation} displays the probability distributions up to $k=5$ calculated for the 4-nearest-neighbors
model (left) and for both the 8-nearest-neighbors and 8-L-connected-neighbors models (right) setting $\varepsilon=0.4$.
The comparison with the distributions of equations (\ref{n_model}-\ref{h_nc}) for $n=4$ and $n=8$, respectively, shows
a redistribution of probabilities due to saturation, probabilities of $k=2$ and $k=3$ being increased while those of
more triggered pixels decreased. This can be interpreted as a consequence of the reduction of the effective number of
active neighbors (see figure~\ref{fig:n_models}). Saturation effects are significant in the 4-nearest-neighbors model,
while they are rather small and weakly dependent on the geometrical arrangement of the neighborhood in both models of 8
neighbors. In fact, saturation is almost negligible in the all-neighbors model for $N\gtrsim100$, which yields results
very similar to the Borel distribution.

The mean, variance and excess noise factor of the probability distributions given in table~\ref{tab:probabilities} were
found to be described, up to second order of approximation in $p$ or $\varepsilon$, by the following unified
expressions:
\begin{equation}
\label{mean_geometrical_models}
E=1+n\,p+n(n-1)p^2+o\left(p^2\right)=1+\varepsilon+\frac{3n-3}{2n}\varepsilon^2+o\left(\varepsilon^2\right)
\end{equation}
\begin{equation}
\label{variance_geometrical_models}
Var=n\,p+n(3n-4)p^2+o\left(p^2\right)=\varepsilon+\frac{7n-9}{2n}\varepsilon^2+o\left(\varepsilon^2\right)
\end{equation}
\begin{equation}
\label{ENF_geometrical_models}
ENF=1+n\,p+n(n-4)p^2+o\left(p^2\right)=1+\varepsilon+\frac{3n-9}{2n}\varepsilon^2+o\left(\varepsilon^2\right)\,,
\end{equation}
only depending on the number of neighbors assumed in each model, i.e., $n=$ 4, 8 or $N-1$. In a third or higher-order
approximation, this is no longer possible since the particular geometrical arrangement of the neighborhood also plays a
role in the development of crosstalk histories. It should be noted that the $E$, $Var$ and $ENF$ parameters for the
proposed geometrical models are lower than those for the models of section~\ref{ssec:fundamentals} for the same number
of neighbors but without saturation effects (see equations (\ref{mean_n_model}-\ref{ENF_n_model})).

We checked that a geometric extrapolation is a good approximation for the probabilities of $k>5$ for the proposed
geometrical models

\begin{equation}
\label{geometric_extrapolation}
P(k)\simeq P(5)\left[1-\frac{P(5)}{1-\sum_{k=1}^{4}P(k)}\right]^{k-5}\,,
\end{equation}
where the geometric rate of decrease is set so that $\sum_{k=1}^{\infty}P(k)=1$. For instance, this allows us to
compute numerically the mean, variance and excess noise factor of the above probability distributions for larger
$\varepsilon$ values and with better precision than equations
(\ref{mean_geometrical_models}-\ref{ENF_geometrical_models}).

\subsection{Photon-counting statistics under pulsed illumination}
\label{ssec:random_primaries}

The above probability distributions are derived assuming a single initially fired pixel and they can be applied to
describe the response of SiPMs with crosstalk at dark conditions or low-intensity illumination. For measurements under
pulsed illumination, simultaneous random excitation of primary pixels (usually Poisson distributed) has to be accounted
for. The composition of a Poisson distribution of initially fired pixels with either a geometric or a Borel
distribution (cases $n=1$ and $n\rightarrow\infty$ in equations (\ref{n_model}-\ref{h_nc})) has already been used by
\cite{Vinogradov1,Vinogradov2}. In a general case, the probability distribution of the total number of triggered
pixels, $k=m$ primaries $+ c$ crosstalk events, is given by

\begin{align}
\label{total}
&P_{\rm tot}(0)=P_{\rm pr}(0) \nonumber\\
&P_{\rm tot}(k)=\sum_{m=1}^{k}P_{\rm pr}(m)\,P_m(k)\,,\quad k=1,\,2,\,3,\ldots\,,
\end{align}
where $P_{\rm pr}(m)$ is the probability distribution of the number $m$ of primary events and $P_m(k)$ is the
probability of total number $k$ of triggered pixels provided $m$ primaries. The latter probability can be calculated
from the recursive formula

\begin{equation}
\label{P_mk}
P_{m+1}(k)=\sum_{i=1}^{k-m}P_m(k-i)\,P_1(i)\,,\quad m=1,\,2,\,3,\ldots\,,
\end{equation}
with $P_1(i)$ being the probability distribution of the number $i$ of triggered pixels for a single initially fired
pixel, like those given in table~\ref{tab:probabilities} for the proposed geometrical models.

These expressions do not account for possible further saturation effects caused by primary events in close pixels
initiating crosstalk histories that overlap each other. Nevertheless, if the number of primary events is low compared
to the total number of pixels of the device, the probability of overlapping crosstalk histories is negligible and this
approximation can be done. In fact, more than 20\% of pixels of the array should be fired initially so that this
saturation effect is important for typical crosstalk probabilities in the 4-nearest-neighbors model \cite{Dovrat}.

In principle, the original photon statistics $P_{\rm pr}(m)$ can be extracted from the observed probability
distribution $P_{\rm tot}(k)$ if the crosstalk distribution $P_1(k)$ is known (e.g., from the pulse-height spectrum of
dark counts). Using (\ref{total}-\ref{P_mk}), the following sequence is derived:

\begin{align}
\label{P_prim}
&P_{\rm pr}(0)=P_{\rm tot}(0) \nonumber\\
&P_{\rm pr}(1)=\frac{P_{\rm tot}(1)}{1-\varepsilon} \nonumber\\
&P_{\rm pr}(m)=\frac{1}{(1-\varepsilon)^m}\left[P_{\rm tot}(m)-\sum_{l=1}^{m-1}P_{\rm pr}(l)\,P_l(m)\right]\,,\quad m=2,\,3,\,4,\ldots\,.
\end{align}
However, this can more easily be accomplished making simple assumptions on the expected photon statistics. In the most
usual case of a Poisson distribution, the photon statistics is entirely determined by the mean, which is simply given
by

\begin{equation}
\label{mu}
\mu=-\ln\left(P_{\rm tot}(0)\right)\,,
\end{equation}
where $P_{\rm tot}(0)$ is unaffected by crosstalk indeed. Once $\mu$ is obtained, the crosstalk probability can be
calculated as

\begin{equation}
\label{epsilon}
\varepsilon=1-\frac{P_{\rm tot}(1)}{\mu\,\exp(-\mu)}\,.
\end{equation}

Note that expressions (\ref{mu}) and (\ref{epsilon}) are independent of the crosstalk model, and we will use them in
section~\ref{sec:results} to calculate respectively the $\mu$ and $\epsilon$ parameters from experimental
data.\footnote{This procedure has previously been employed in \cite{Vacheret}.}. In other cases, however, it may not be
possible to use this approach either because the photon arrival is not Poisson distributed or because the $P_{\rm
tot}(0)$ and $P_{\rm tot}(1)$ probabilities cannot be determined accurately (e.g., if the number of impinging photons
is large, these probabilities will be too small).

From (\ref{total}-\ref{P_mk}), it can be demonstrated that the mean and variance of the observed probability
distribution are related to those of photons and crosstalk events as

\begin{equation}
\label{mean_total}
E_{\rm tot}=E_{\rm pr}\,E_1
\end{equation}
\begin{equation}
\label{variance_total}
Var_{\rm tot}=E_{\rm pr}\,Var_1+Var_{\rm pr}\,E^2_1\,,
\end{equation}
where $E_1$ and $Var_1$ can be calculated using the approximate expressions given in table~\ref{tab:probabilities} for
the proposed crosstalk models.\footnote{In this paper, we use a simplified notation in which the names of the random
variables are omitted (symbols for their numerical values are only shown), and indexes are used instead to indicate the
corresponding random variable. For instance, $P_{\rm tot}(k)$, $E_{\rm tot}$ and $Var_{\rm tot}$ stand respectively for
$P(K_{\rm tot}=k)$, $E[K_{\rm tot}]$ and $Var[K_{\rm tot}]$, where $K_{\rm tot}$ is the random variable associated to
these probability distribution, mean and variance.} Therefore, valuable information of the original probability
distribution of photons can be obtained simply from the mean and variance of the measured pulse-height spectrum if the
crosstalk probability is known.

To evaluate the excess noise introduced by crosstalk, the contribution of the random excitation of primary pixels
(i.e., input noise) should be discounted from the total noise in the output signal. Following
\cite{Vinogradov1,Vinogradov2}, the excess noise factor due to crosstalk can be defined as the relative losses in the
signal-to-noise ratio from input to output

\begin{equation}
\label{ENF2}
ENF'=\frac{E^2_{\rm pr}/Var_{\rm pr}}{E^2_{\rm tot}/Var_{\rm tot}}\,.
\end{equation}
However, this definition is not equivalent to the standard one given by (\ref{ENF}) for dark-count measurements. We
found that the standard definition can be generalized for pulsed-illumination conditions using the following
alternative expression:

\begin{equation}
\label{ENF_general}
ENF=1+E_{\rm pr}\left(\frac{Var_{\rm tot}}{E^2_{\rm tot}}-\frac{Var_{\rm pr}}{E^2_{\rm pr}}\right)
=1+\frac{Var_1}{E_1^2}\,.
\end{equation}
Note that when $E_{\rm pr}=Var_{\rm pr}$, as happens for a Poisson distribution, both expressions (\ref{ENF2}) and
(\ref{ENF_general}) are identical, whereas only (\ref{ENF_general}) properly reduces to (\ref{ENF}) when $E_{\rm pr}=1$
and $Var_{\rm pr}=0$ (e.g., dark counts with a single primary event).

\section{Experimental method}
\label{sec:experiment}

A dedicated experiment was carried out to study the crosstalk effect in a SiPM device. Two types of measurements were
performed: under low-light-level conditions (dark counts) and using a pulsed light source to illuminate the device.

The experimental setup is described in section~\ref{ssec:setup}. The data analysis, which includes the processing of
signals with a specific software, is explained in section~\ref{ssec:analysis}.

\subsection{Setup}
\label{ssec:setup}

\begin{figure}[t]
\begin{center}
\includegraphics[width=.8\textwidth]{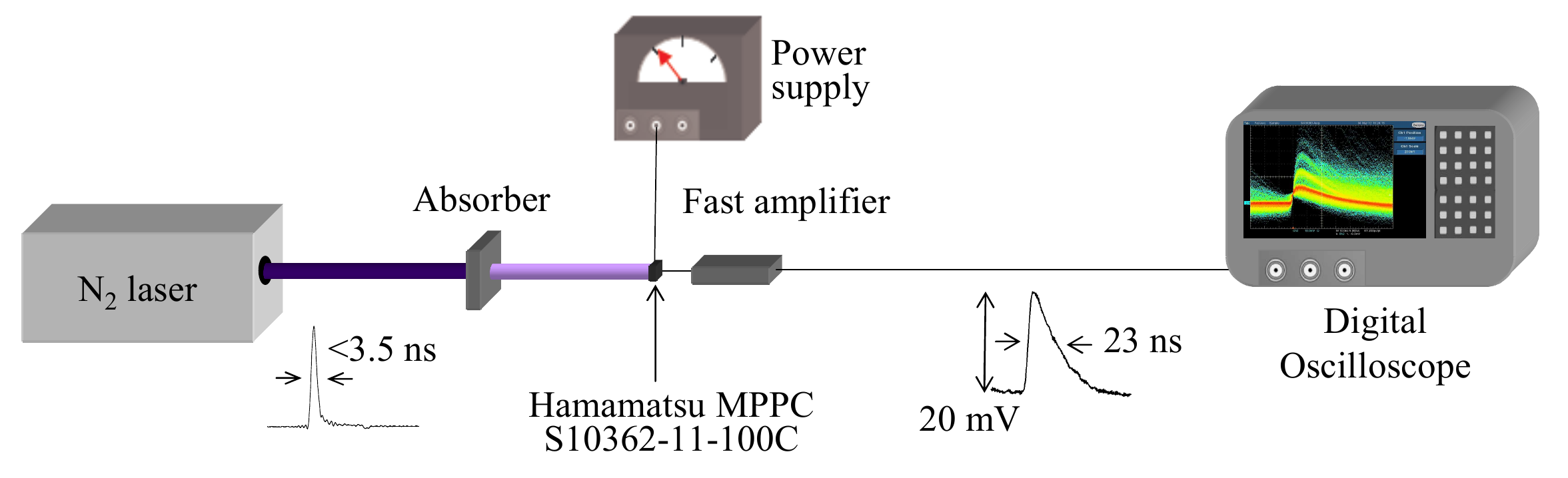}
\end{center}
\caption{Schematic view of the experimental setup consisting of a SiPM (Hamamatsu S10362-11-100C) with associated electronics,
a N$_2$ laser emitting 3.5~ns FWHM pulses and a digital oscilloscope.}
\label{fig:Setup}
\end{figure}

A schematic view of the experimental setup is shown in figure~\ref{fig:Setup}. This includes the SiPM with the
associated electronics, the light source and the data-acquisition system. The SiPM chosen for this study was the model
S10362-11-100C from Hamamatsu with 100 pixels in an active area of 1~mm $\times$ 1~mm. It was connected to a bias
circuit that had a load resistor of 1~k$\Omega$ and a coupling capacitor of 0.1~$\mu$F. Signals were sent to a fast
amplifier based on AD8367 (Analog Devices) with a bandwidth of 500~MHz and nominal gain of 42.5~dB. The amplifier was
configured to provide narrow output pulses with a rise time of about 2~ns and an exponential decay with 23~ns time
constant. Typical pulse sizes of 20~mV were achieved for single-photoelectron signals when working at an overvoltage of
around 1.2~V.

For the measurements under pulsed illumination, we used a 337~nm nitrogen laser (Stanford Research Systems, model
NL100) emitting light pulses of 3.5~ns FWHM at a repetition rate of up to 20~Hz. The laser beam was attenuated and
arranged to illuminate uniformly the active area of the detector. The laser-pulse intensity was monitored using an
internally installed photodiode, which also provided the synchronized output signal.

Data were recorded using a digital oscilloscope (Tektronix TDS5032B) with a bandwidth of 350~MHz and a sampling rate of
up to 5~GS/s. The oscilloscope enabled us to store all the time-intensity information of signals in ASCII files for
later analysis. In dark-count measurements, each file contained a 400~$\mu$s long run, while in pulsed regime, the data
acquisition was done within a time window of 200~ns in coincidence with the laser pulse (4000 laser shots per file).
The time resolution was set to 0.4~ns per channel in both cases.

Both bias voltage and temperature were monitored. The bias voltage was stable within 0.1\% during each run. Temperature
was measured using a type K thermocouple close to the SiPM, enabling us to register changes of 0.5~$^\circ$C.

\subsection{Waveform analysis}
\label{ssec:analysis}

\begin{figure}[t]
\begin{center}
\includegraphics[width=.4\textwidth]{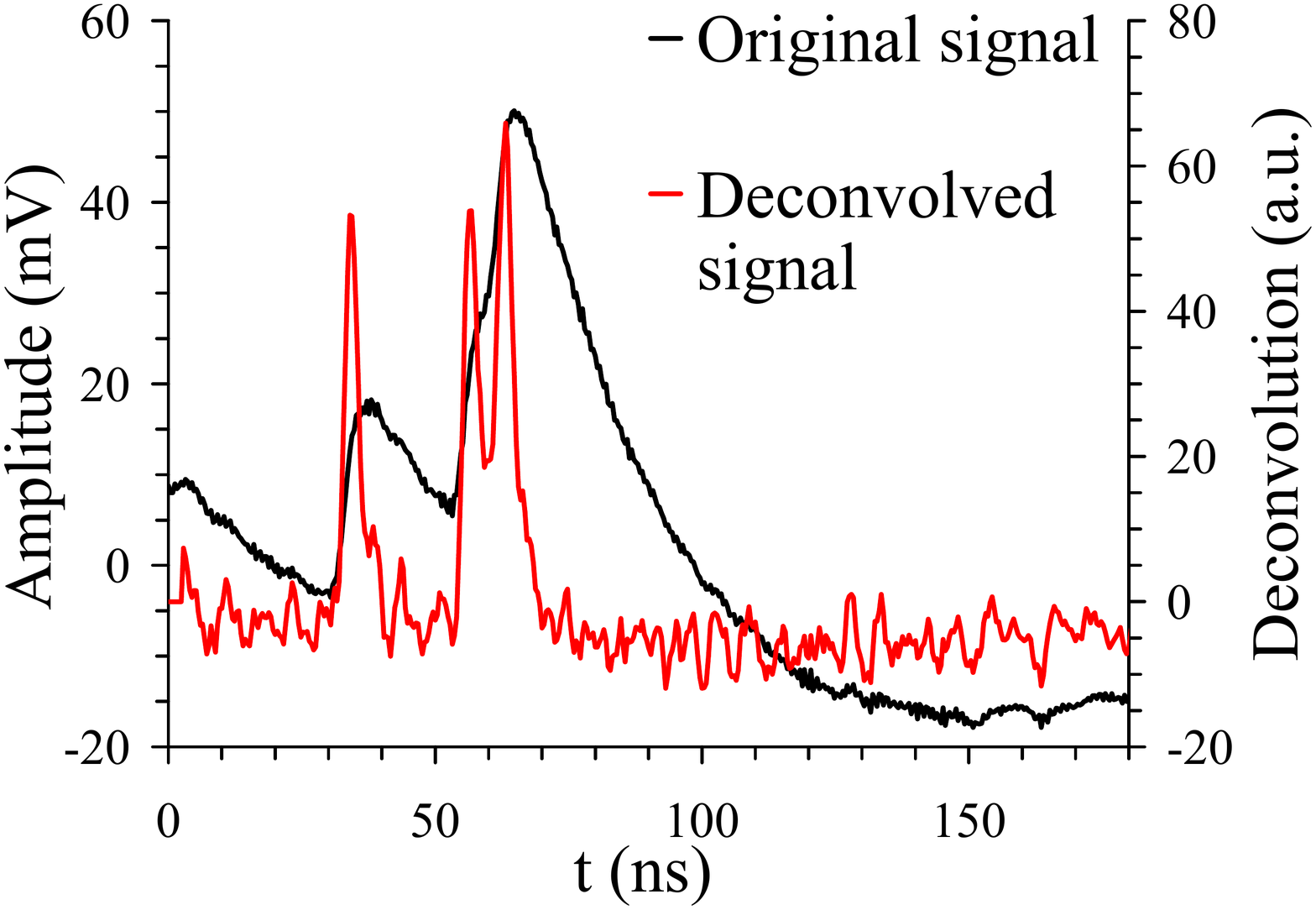}
\includegraphics[width=.4\textwidth]{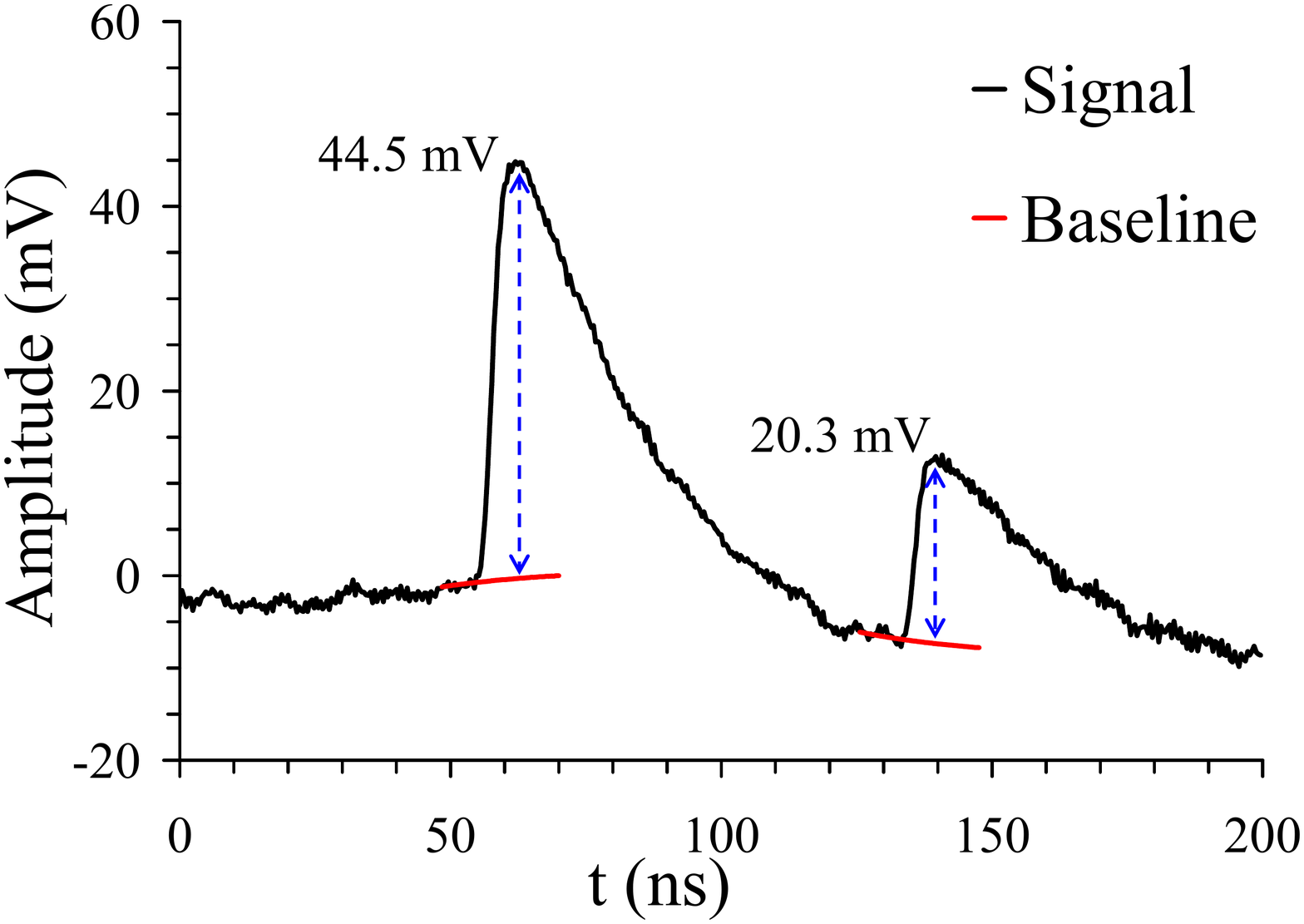}
\end{center}
\caption{Dark counts registered at an overvoltage of 1.2~V. Left: Identification of pulses through the signal deconvolution.
Right: Determination of the pulse amplitude using an exponential fit to determine the baseline of each pulse.}
\label{fig:pulses}
\end{figure}

A C++ program was developed to analyze the raw data stored by the digital oscilloscope. Essentially, the algorithm
searched for pulses and then measured their arrival times and amplitudes. A linear digital filtering of the signal was
also performed to reduce the electronic noise. Some sample pulses registered at dark conditions are shown in
figure~\ref{fig:pulses} (the digital filter is already applied to signals).

Pulses were identified by their sharp leading edge. To this end, the signal was previously deconvolved with an
exponential decay function with 23~ns time constant, in such a way that tails of pulses were suppressed and edges were
clearly distinguished even for piled-up pulses (left-hand plot of figure~\ref{fig:pulses}). Threshold criteria were
applied to this deconvolved signal on both amplitude and pulse width to further discriminate electronic noise. Then,
the arrival time of an identified pulse was measured as the time at which the deconvolved signal reaches the maximum,
corresponding to the position of the pulse edge in the original signal. The smallest time difference to resolve close
pulses was determined to be $3.0\pm0.6$~ns, which was our coincidence time window in dark-count measurements (crosstalk
is assumed to be nearly instantaneous). On the other hand, when illuminating with the laser beam, light was detected
within a time interval of 5~ns, which was selected as the coincidence window instead, that is, two pulses that arrive
within this time interval are considered as a single pulse even though they can actually be resolved.

The pulse amplitude was measured as the peak height from the baseline. The original signal was used for this purpose,
because using the deconvolved signal resulted in larger uncertainties. Since pulses are often on the tail of the
preceding one (see figure~\ref{fig:pulses}), the baseline of each individual pulse was calculated by fitting the function

\begin{equation}
\label{exponential_fit}
f(t)=A+B\,\exp\left(-\frac{t}{\tau}\right)\,,\quad {\rm with}\;\tau\equiv23\:{\rm ns}\,,
\end{equation}
to data in the time interval of 8~ns prior to the pulse, and extrapolating it up to the peak time. This method was
proved to significantly enhance the measured pulse-height resolution of both dark counts and laser pulses. For
instance, the amplitudes of the pulses of the right-hand plot of figure~\ref{fig:pulses} were determined to be 44.5~mV
and 20.3~mV, which are typical amplitudes of pulses having two and one triggered pixels, respectively, at an
overvoltage of 1.2~V. However, limitations were found for piled-up pulses. If two pulses are closer than 20~ns, the
amplitude of the second one cannot be accurately calculated because the fit of equation (\ref{exponential_fit}) to data
fails. In addition, if the time difference is less than 10~ns, the amplitude of the first pulse cannot be measured
either, because the peak is not well distinguished from the second pulse. Therefore, although the piled-up pulses of
the left-hand plot of figure~\ref{fig:pulses} are resolved (time difference of 6.4~ns), their amplitudes would not be
measured separately. Quality cuts were applied to the identified pulses to avoid these problems, resulting in a further
improvement of the pulse-height resolution. For the evaluation of crosstalk, a dead time of 120~ns was also introduced
after each identified pulse to prevent afterpulsing effects (see section~\ref{ssec:afterpulsing}).

Our analysis software generated a table of arrival times and amplitudes of the identified pulses classified according
to the quality cuts and dead-time criteria. This allowed us to obtain pulse-height spectra and to study possible
time-amplitude correlations as well as the effect of the applied cuts. The algorithm also calculated the rate of pulses
in dark-count measurements. For this purpose, the acquisition time was calculated taking into account the applied dead
time and quality cuts.

\section{Results}
\label{sec:results}

Experimental results are presented and discussed below. In the first place, amplitude and timing measurements of
afterpulsing, relevant for a correct treatment of experimental data, are reported. In section~\ref{ssec:dark_counts},
the probability distribution of the number of triggered pixels is extracted from the pulse-height spectrum measured at
dark-count conditions and it is used to discriminate between the crosstalk models described above. In
section~\ref{ssec:pulsed_illumination}, applicability of equations (\ref{mean_total}) and (\ref{variance_total}) to
describe the effects of crosstalk on some pulse-height spectra measured under pulsed-illumination conditions is
evaluated.

\subsection{Afterpulsing and recovery time}
\label{ssec:afterpulsing}

Parasitic avalanches can be produced by the delayed release of carriers trapped in deep-level defects during a previous
avalanche \cite{Haitz,Cova}. This effect is referred to as afterpulsing and, unlike crosstalk, it occurs in the same
pixel where the primary avalanche was developed. After breakdown, a pixel need some time to recover the original bias
voltage, so secondary avalanches with a delay less than the recovery time are weaker than ordinary ones. Short-delay
afterpulses usually appear as small humps on the tail of the primary pulse and do not affect the pulse-height
measurement. However, they can be counted as separate pulses of lower amplitude, distorting the pulse-height spectrum.
Furthermore, afterpulsing avalanches are obviously able to induce crosstalk, but low-intensity ones are expected to
have a smaller probability, and thus, they would contribute to reduce the average crosstalk probability if they are not
discriminated. As pointed out in the previous section, our analysis algorithm applies a dead time long enough to
prevent these effects due to afterpulse contamination.

\begin{figure}
\begin{center}
\includegraphics[width=.8\textwidth]{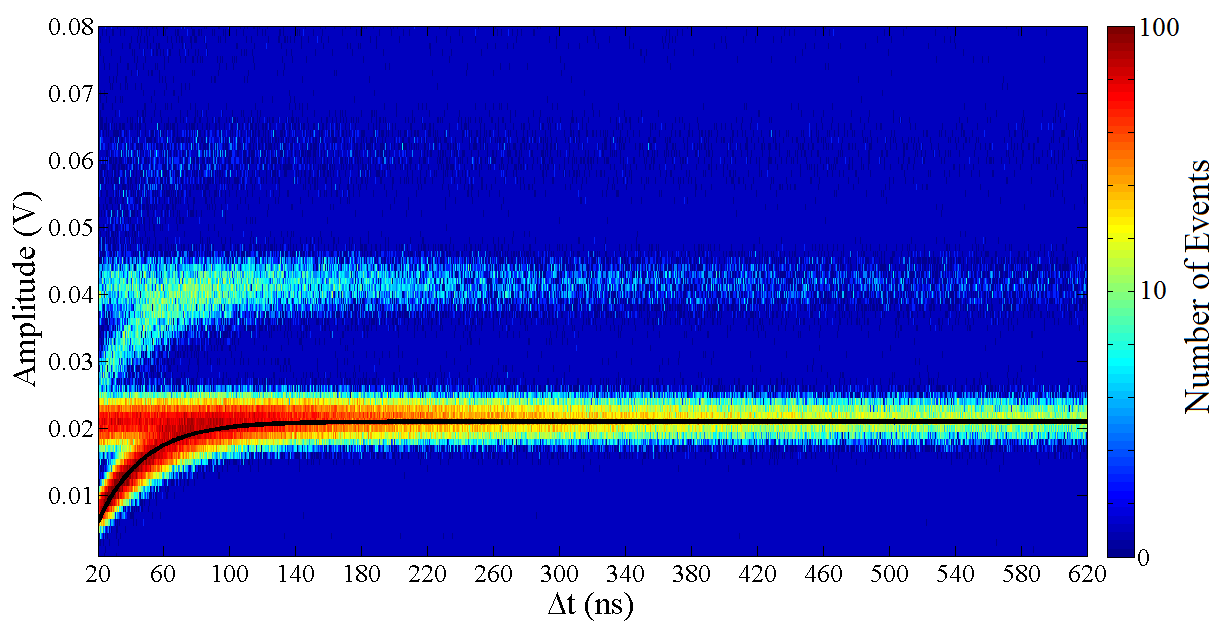}
\end{center}
\caption{Two-dimensional histogram of the pulse amplitude versus the arrival time relative to the preceding pulse.
Two branches can be seen due to contributions of non-correlated events and afterpulses, respectively.
An exponential fit to the lower branch is shown, providing an estimate of the recovery time.
Crosstalk effects result in the repetition of the same pattern of branches at higher amplitudes.}
\label{fig:afterpulsing}
\end{figure}

The recovery time of our SiPM was estimated using a technique similar to \cite{Oide,Du}, that is, studying the
dependence of the afterpulse amplitude on the delay relative to the primary pulse. To this end, dark counts were
registered continuously in a wide time window and analyzed without imposing dead time so that afterpulses were also
included. Given any two consecutive pulses, they may be either uncorrelated events or an afterpulse following its
primary event. Uncorrelated avalanches are generally induced in fully recharged pixels, and thus, they should have
constant amplitude, while afterpulses have an amplitude proportional to the overvoltage reached in the pixel since the
last avalanche breakdown, with an expected exponential recovery. Both contributions are clearly visible in the
two-dimensional histogram displayed in figure~\ref{fig:afterpulsing}, showing that pulse amplitudes are distributed in
a narrow band that splits into two branches when the time difference with respect to the preceding pulse is smaller
than $\sim80$~ns, the lower branch being due to short-delay afterpulses. To better resolve these contributions, pulses
closer than 80~ns to their two preceding pulses were rejected, that is, potential short-delay afterpulses that are not
produced by the immediately preceding avalanche (fired in a different pixel) were eliminated. Crosstalk effects result
in the repetition of the same pattern with two branches at higher amplitudes. The cut at 20~ns is due to the
limitations of our analysis algorithm to calculate the amplitude of piled-up pulses (section~\ref{ssec:analysis}).

As expected, the afterpulse amplitude grows exponentially at increasing delay, although the recovery curve was found to
be more suitably described if an offset time of 5~ns is added. We measured a recovery time (i.e., the reciprocal of the
exponential rate) of 40~ns with an estimated uncertainty of 5~ns, which is consistent with the result reported in
\cite{Oide} for the same SiPM device and at room temperature.

\subsection{Dark-count measurements}
\label{ssec:dark_counts}

\begin{figure}[t]
\begin{center}
\includegraphics[width=.6\textwidth]{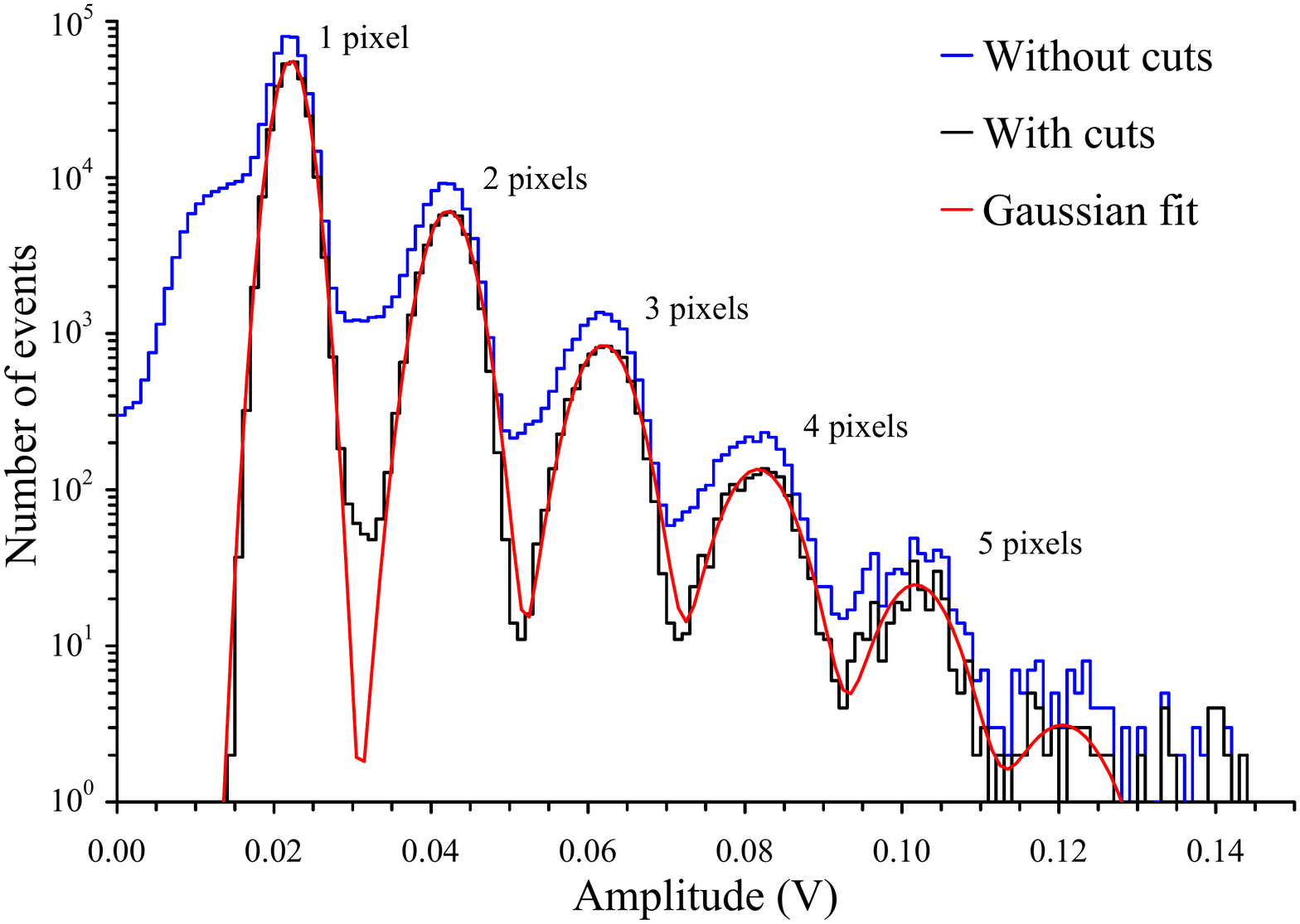}
\end{center}
\caption{Pulse-height spectrum of dark counts measured at an overvoltage of 1.2~V and room temperature.
The applied cuts and the high statistics of events have allowed to clearly resolved five peaks.
A Gaussian fit was used to disentangle contributions of adjoining peaks.}
\label{fig:dark_counts}
\end{figure}

The pulse-height spectrum of dark counts was measured at an overvoltage of 1.2~V and room temperature (25~$^\circ$C).
In figure~\ref{fig:dark_counts}, the experimental spectrum is shown with and without applying the cuts described in
section~\ref{ssec:analysis}. The controlled dead time introduced in the data analysis reduces significantly the
statistics, while it also removes the pronounced shoulder on the left of each peak due to low-amplitude afterpulses
(see section~\ref{ssec:afterpulsing}). Quality cuts discriminate a very small fraction of piled-up pulses that mostly
populate the valleys, resulting in a slight improvement of the height-pulse resolution with no significant bias in the
experimental probability distribution. Five Gaussian-like peaks were clearly resolved in the spectrum corresponding to
pulses from 1 to 5 triggered pixels, although pulses of higher amplitudes were also registered. The number of collected
events passing the cuts was $3.1\cdot10^5$.

The probability that a pulse has $k=1$, 2, 3, 4, 5 triggered pixels, $P(k)$, was determined as the area under the
$k$-th peak divided by the total number of events in the spectrum. Small corrections were applied to disentangle
contributions of adjoining peaks in the valley region by means of Gaussian fitting. However, peaks are not exactly
Gaussian and, consequently, differences between the experimental peak areas and those of the fitted Gaussian curves
were added to the uncertainties, which are still dominated by statistical errors.

In dark-count measurements, the crosstalk probability $\varepsilon$ can be estimated as the fraction of pulses with two
or more triggered pixels, i.e., $1-P(1)$, which was measured to be $0.1592\pm0.0007$. Accidental coincidences of
non-correlated dark-counts events, however, would contribute to this fraction even in the absence of crosstalk. The
observed dark-count rate, excluding afterpulses, was 1.76~MHz, and therefore, the expected average number of events
within a time window of $3.0\pm 0.6$~ns (see section~\ref{ssec:analysis}) is $\mu=0.0053\pm 0.0011$. Assuming a Poisson
distribution of dark counts, the probability that a given pulse is due to two or more coincident dark-count events is
only 0.0026. To account for this small contribution of accidental coincidences, the crosstalk probability was
determined using (\ref{epsilon}), resulting in $\varepsilon=0.1570\pm 0.0008$. Note that $P_{\rm tot}(k)=(1-P_{\rm
tot}(0))\,P(k)$ for $k\geq1$ has to be used for application of equation (\ref{epsilon}), where $P_{\rm
tot}(0)=\exp(-\mu)$.

\begin{figure}[t]
\begin{center}
\includegraphics[width=.4\textwidth]{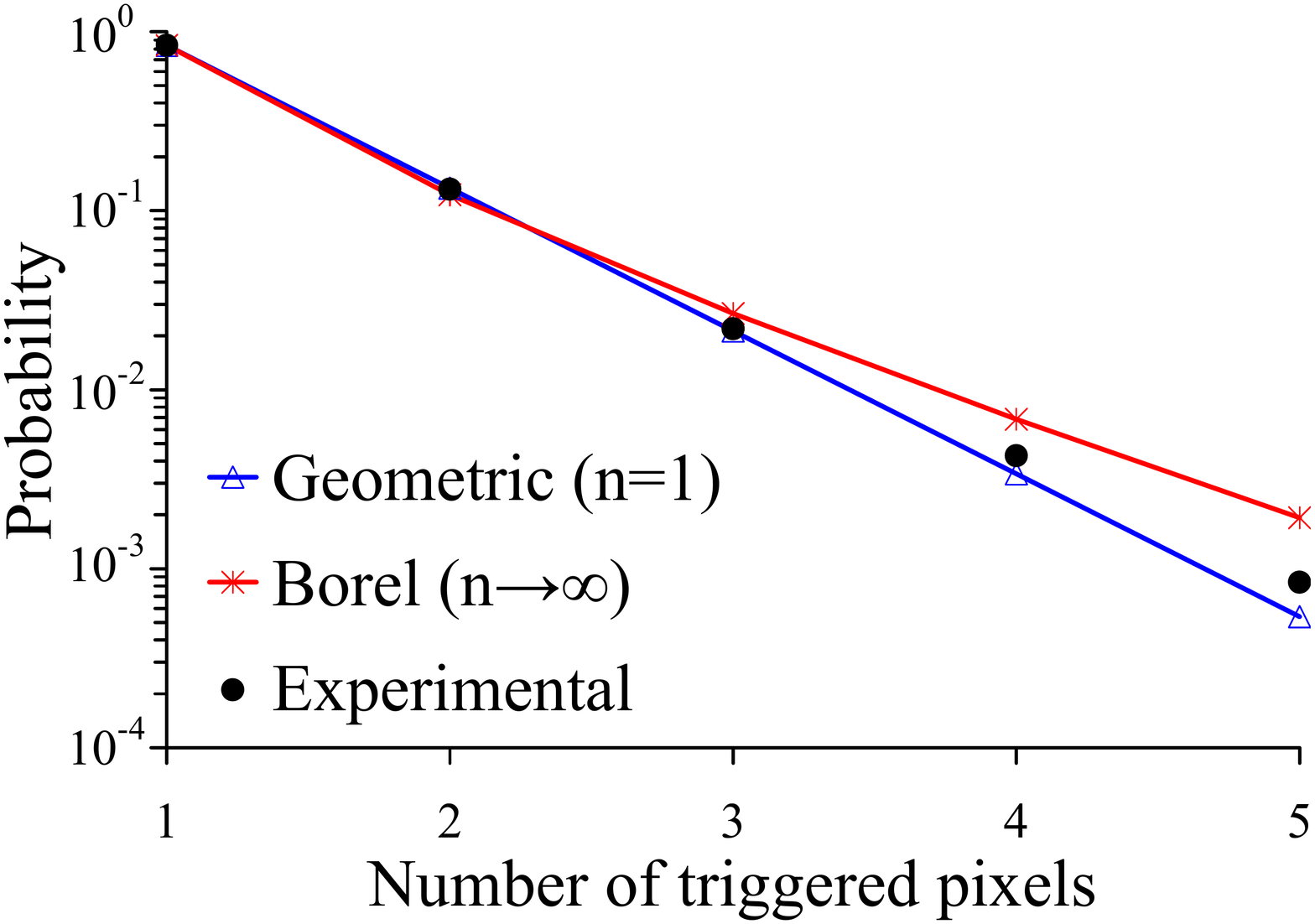}
\includegraphics[width=.4\textwidth]{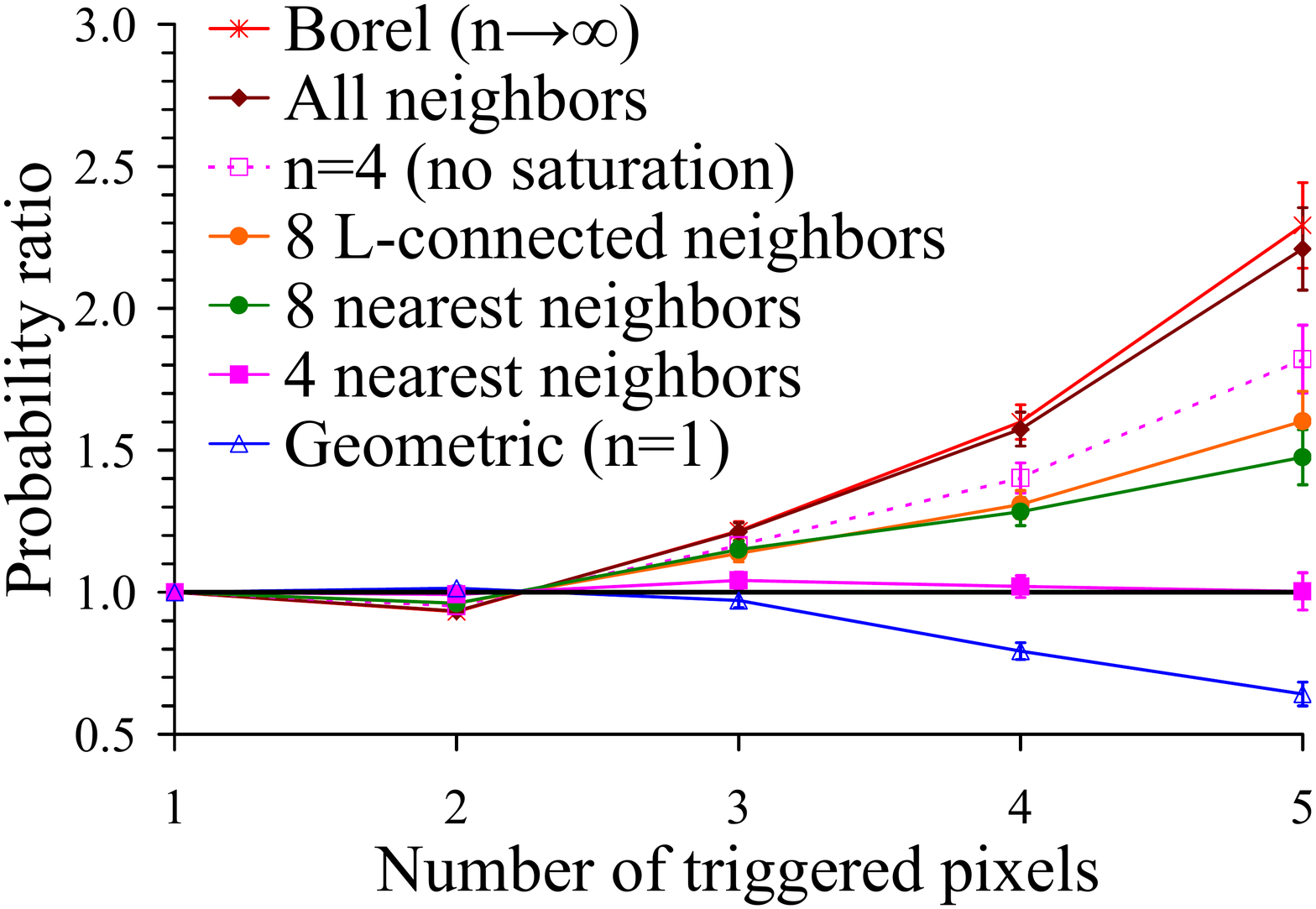}
\end{center}
\caption{Comparison of experimental probabilities with theoretical predictions from different crosstalk models.
Left: Experimental data lie between the two limit situations corresponding to both geometric and Borel distributions
of crosstalk~\cite{Vinogradov1,Vinogradov2}.
Right: Ratio of theoretical and experimental probabilities for different crosstalk models.
Only the 4-nearest-neighbors model is compatible with experimental data.}
\label{fig:comparison_dark_counts}
\end{figure}

Theoretical probability distributions were calculated for the different crosstalk models explained above. For this
purpose, equations (\ref{total}-\ref{P_mk}) were used setting $\varepsilon=0.1570$ and assuming that $P_{\rm pr}(m)$ is
a Poisson distribution with mean $\mu=0.0053$. In the left-hand plot of figure~\ref{fig:comparison_dark_counts},
experimental probabilities are compared with those predicted by the crosstalk models of~\cite{Vinogradov1}
and~\cite{Vinogradov2}, based on the geometric and Borel distributions, respectively. The experimental data lie between
these two limit situations, as expected. A geometric distribution ($n=1$) underestimates the probability of multiple
crosstalk events, while the Borel distribution ($n\rightarrow\infty$) overestimates it. In the right-hand plot, the
ratio of theoretical and experimental probabilities are displayed for these two crosstalk models as well as for the
geometrical models described in section~\ref{ssec:saturation}. Error bars represent the experimental uncertainties
including those of the $\epsilon$ and $\mu$ parameters used to calculate the theoretical probabilities. The only model
that is compatible with the experimental data within uncertainties is the 4-nearest-neighbors model. Results of the
model with $n=4$ without saturation effects (section~\ref{ssec:fundamentals}), also included in the figure, deviate
significantly from experimental data, showing that these effects are important to describe the crosstalk statistics.

The excess noise factor was calculated using (\ref{ENF_general}), resulting in a value of $1.1665\pm0.0022$.
Theoretical predictions of the excess noise factor for $\varepsilon=0.1570$ range from 1.1570 (geometric distribution)
to 1.2059 (Borel distribution). Again, the only crosstalk model that gives an excess noise factor compatible with the
experimental one is the 4-nearest-neighbors model ($ENF=1.1683$ is obtained numerically using the extrapolation of
equation (\ref{geometric_extrapolation})). The comparison for the $E_1$ and $Var_1$ parameters yields to the same
conclusion.

Our results confirm the 4-nearest-neighbors hypothesis of crosstalk, previously used in Monte Carlo simulations
\cite{SanchezMajos,Vacheret,Dovrat} to describe different SiPM devices from Photonique and Hamamatsu. This hypothesis
is further reinforced by some experiments with various SiPMs from Hamamatsu and SensL, performing a selective pixel
illumination by means of a microscopic scan with a point light source \cite{Vacheret,Eckert}. The crosstalk probability
was found to be nearly constant when illuminating inner pixels, whereas it is significantly reduced for pixels right at
the edges of the array, since they have fewer neighbors.

It should be emphasized that these results are not in disagreement with those of \cite{Rech,Otte}, where crosstalk was
found to be also induced efficiently in distant pixels in some non-commercial SiPM devices. In those cases, crosstalk
may be better described by a wider neighborhood of pixels. In fact, it was shown in \cite{Vinogradov2} that the
experimental pulse-height distribution of dark counts obtained by~\cite{Buzhan}, using their own SiPM design of 1600
pixels, fit well a Borel distribution, which is nearly equivalent to that predicted by our all-neighbors model.
Moreover, crosstalk events with a delay of a few tens of nanoseconds were observed in~\cite{Buzhan}, which were
attributed to a slower crosstalk process where charge carriers are generated in the Si bulk and migrate into the
avalanche region of a perhaps distant pixel (see also~\cite{Otte}). This contribution is disregarded in our analysis,
since it only includes simultaneous excitation of pixels within a 3~ns time window.

\subsection{Measurements under pulsed illumination}
\label{ssec:pulsed_illumination}

\begin{figure}[t]
\begin{center}
\includegraphics[width=.32\textwidth]{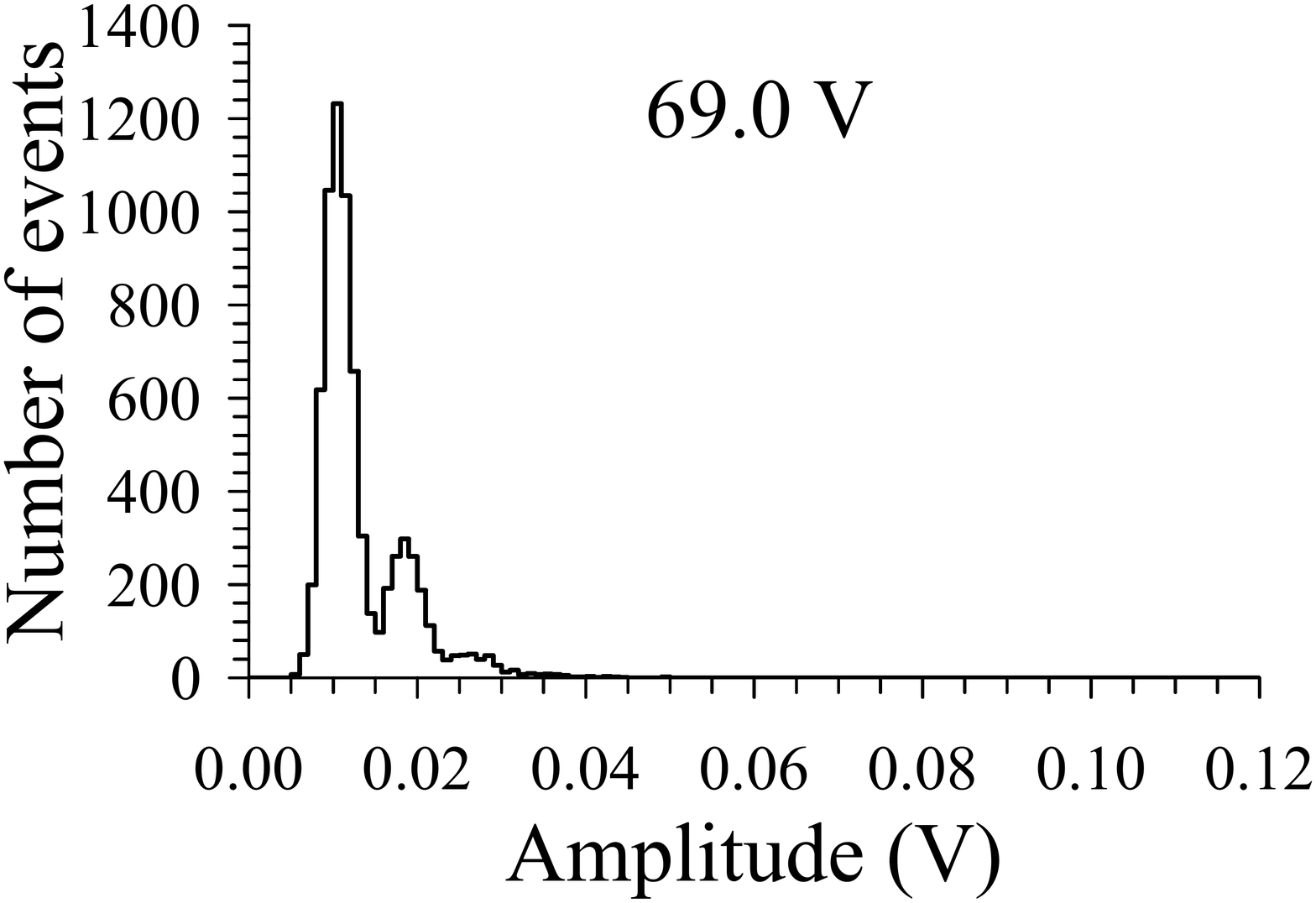}
\includegraphics[width=.32\textwidth]{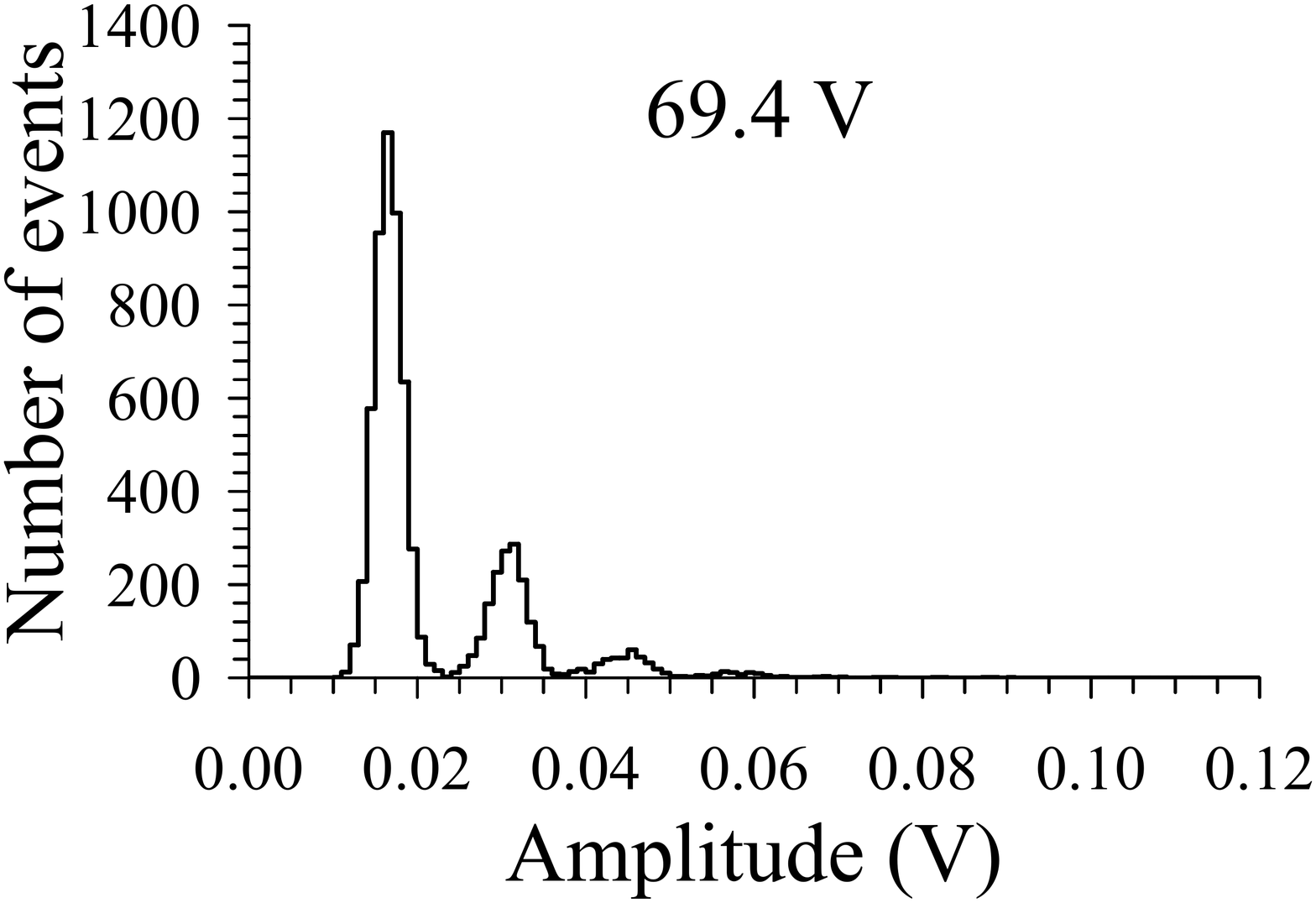}
\includegraphics[width=.32\textwidth]{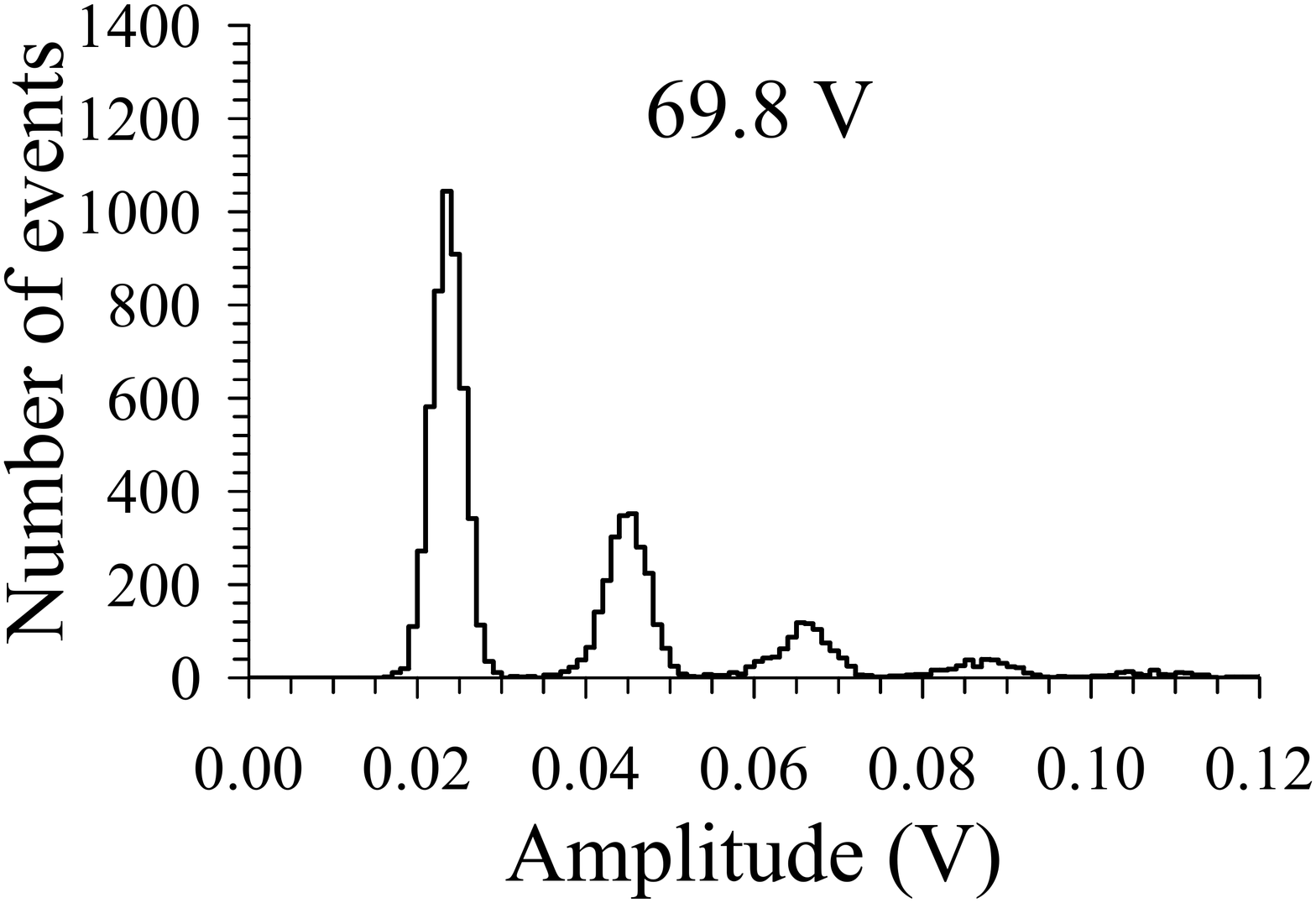}
\end{center}
\caption{Pulse-height spectra measured under pulsed illumination at constant light intensity but different bias voltages.}
\label{fig:pulsed_illumination}
\end{figure}

\begin{figure}[t]
\begin{center}
\includegraphics[width=.32\textwidth]{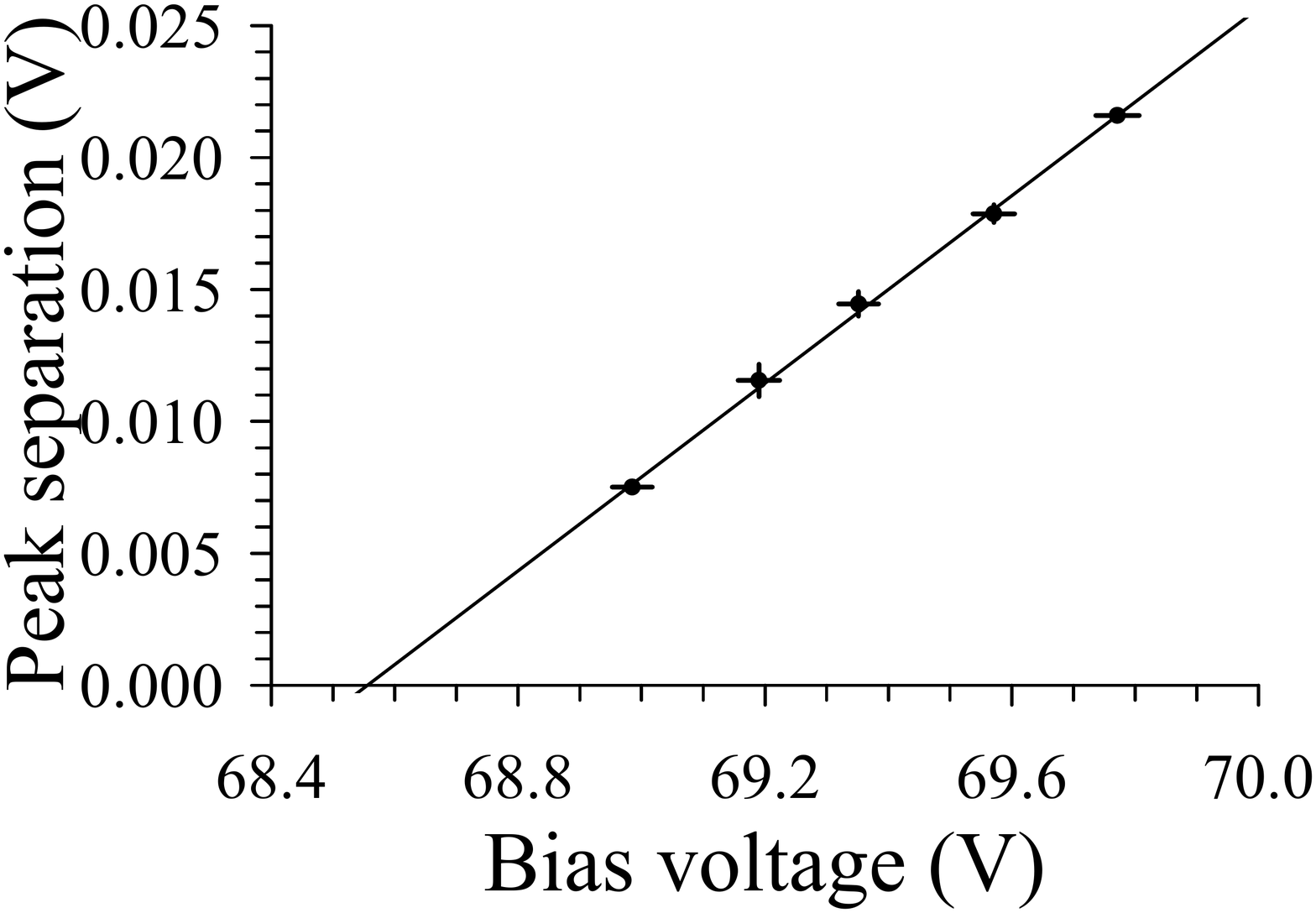}
\includegraphics[width=.32\textwidth]{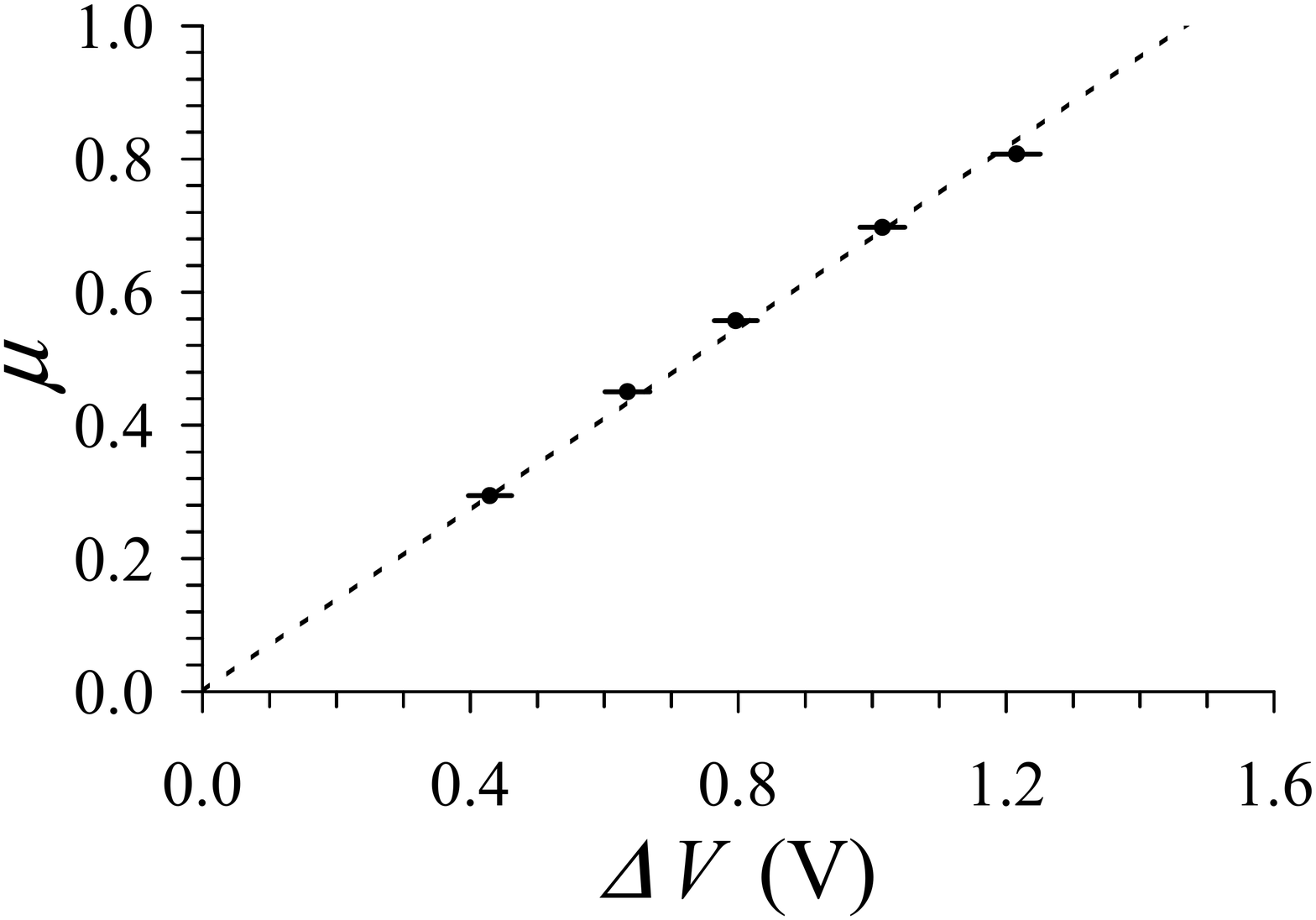}
\includegraphics[width=.32\textwidth]{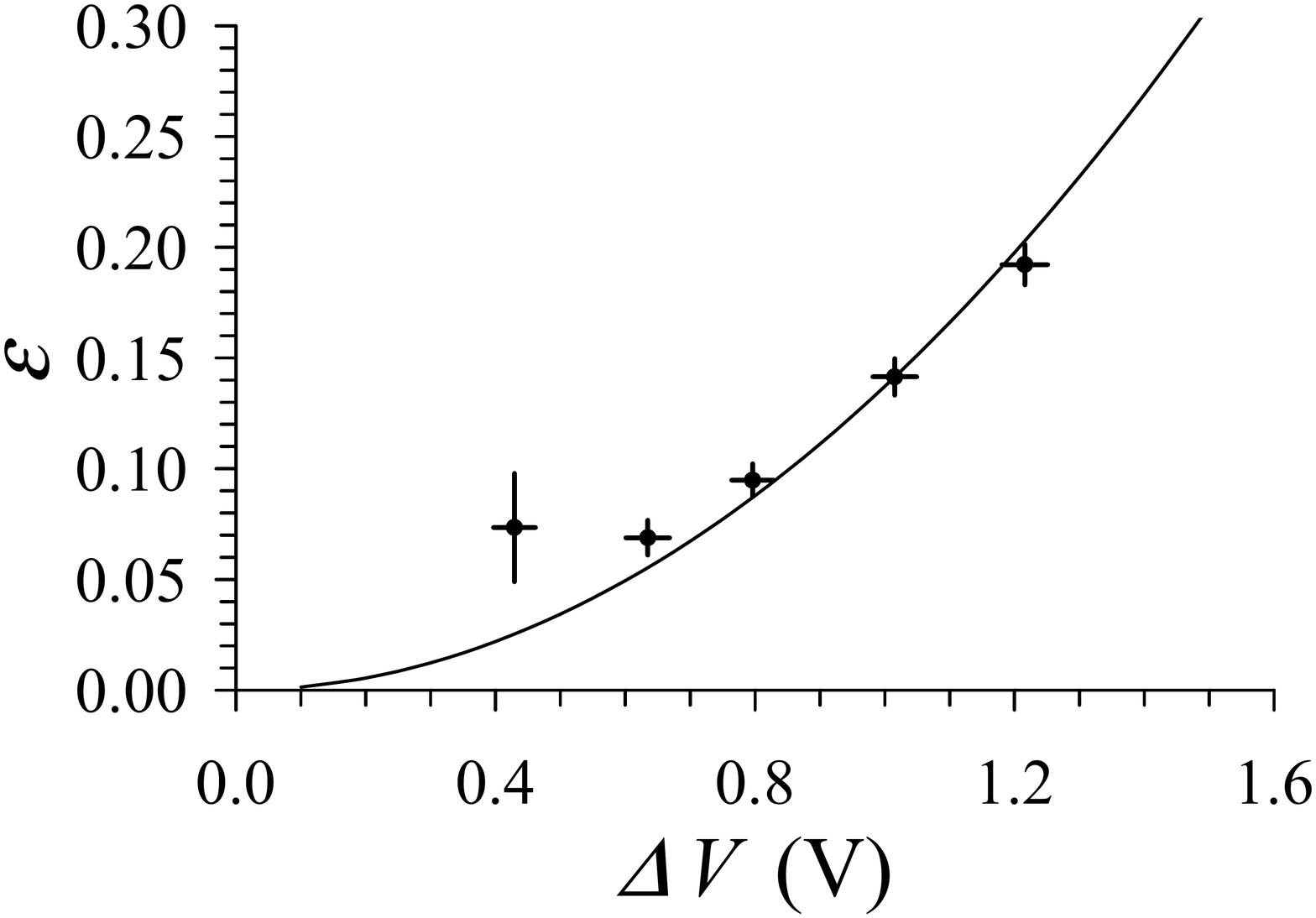}
\end{center}
\caption{
Results of the measured pulse-height spectra. Left: Distance between peaks (proportional to gain) versus bias voltage
and linear extrapolation to obtain the breakdown voltage.
Center: Average number of triggered pixels (proportional to sensitivity) versus overvoltage.
Right: Probability of crosstalk versus overvoltage and fit of equation (\protect\ref{epsilon_overvoltage}) to data.}
\label{fig:results_pulsed_illumination}
\end{figure}

Theoretical relationships deduced in section~\ref{ssec:random_primaries} were validated against experimental data. For
this purpose, a set of measurements was carried out illuminating the SiPM with the pulsed nitrogen laser at constant
conditions (light intensity and temperature) but different SiPM bias voltage. Three pulse-height spectra recorded at
69.0~V, 69.4~V and 69.8~V are shown in figure~\ref{fig:pulsed_illumination}. Avalanche gain is proportional to
overvoltage, hence pulse amplitudes increase linearly at increasing bias voltage, which results in the stretching of
the spectrum. This is illustrated in the left-hand plot of figure~\ref{fig:results_pulsed_illumination}, where the
distance between peaks (proportional to gain) is represented as a function of bias voltage. Data exhibit a pure linear
behavior, as expected, and the extrapolation at zero gain provides the breakdown voltage, which is determined to be
$V_{\rm b}=68.54\pm0.04_{\rm stat}\pm0.79_{\rm syst}$~V at $25^\circ$C.

Measurements were performed at very low light intensity ($<1$~photon detected per laser shot), and a Poissonian photon
statistics was assumed. Therefore, equation (\ref{mu}) allowed us to calculate the actual average number $\mu$ of
detected photons from the fraction of laser shots in which no pulse was registered, i.e., $P_{\rm tot}(0)$.
Contribution of dark counts in coincidence with the laser shot (time window of 5~ns) was estimated to be less than 1\%.
Results are shown in the central plot of figure~\ref{fig:results_pulsed_illumination} as a function of the overvoltage
$\Delta V$, where data were previously normalized by the relative mean intensity of the laser beam measured with the
photodiode during each run to compensate the laser-intensity drift. An approximated linear behavior is found,
indicating that the SiPM sensitivity is essentially proportional to overvoltage (see \cite{Eckert} for a precise
measurement of the sensitivity of the Hamamatsu model S10362-11-100C).

The crosstalk probability was calculated using equation (\ref{epsilon}), where $P_{\rm tot}(1)$ is determined as
described in section~\ref{ssec:dark_counts}. Results exhibit a quadratic dependence on overvoltage in the experimental
range (right-hand plot of figure~\ref{fig:results_pulsed_illumination}), which is expected from the linear growths of
both sensitivity and gain. However, saturation should be reached at large overvoltage. Assuming that the number of
crosstalk photons impinging a given neighbor of the primary pixel is Poisson distributed, the crosstalk probability
should behave as

\begin{equation}
\label{epsilon_overvoltage}
\varepsilon=1-\exp\left(-K\,\Delta V^2\right)\,,
\end{equation}
where $K$ is a constant of the detector. The best fit of this equation to data is also shown in the figure. The
relatively large deviations of data at 69.0~V and 69.2~V from the fitted curve may be due to a lower signal-to-noise
ratio introducing some bias in the calculated $P_{\rm tot}(0)$ and $P_{\rm tot}(1)$ probabilities.

\begin{figure}[t]
\begin{center}
\includegraphics[width=.4\textwidth]{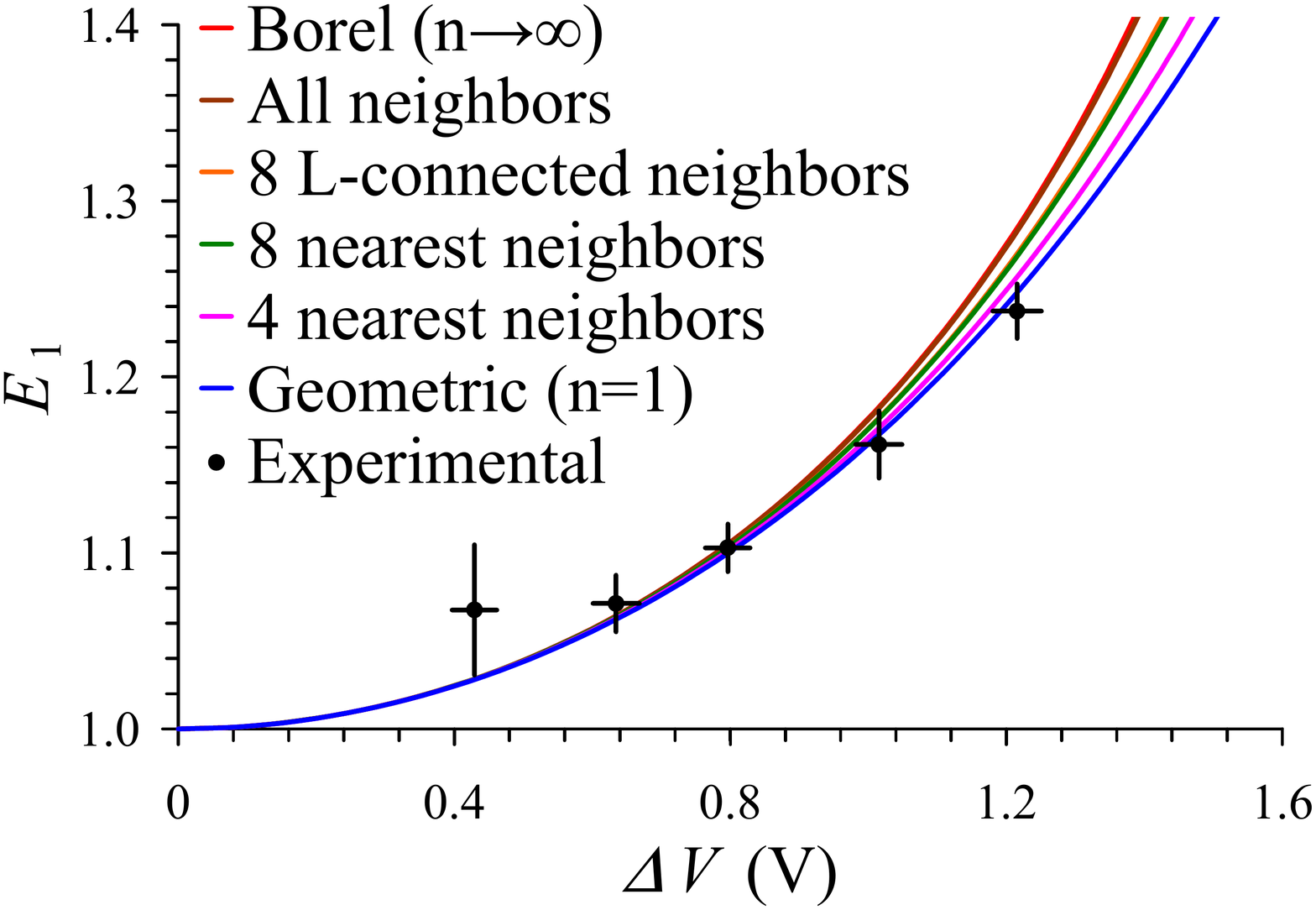}
\includegraphics[width=.4\textwidth]{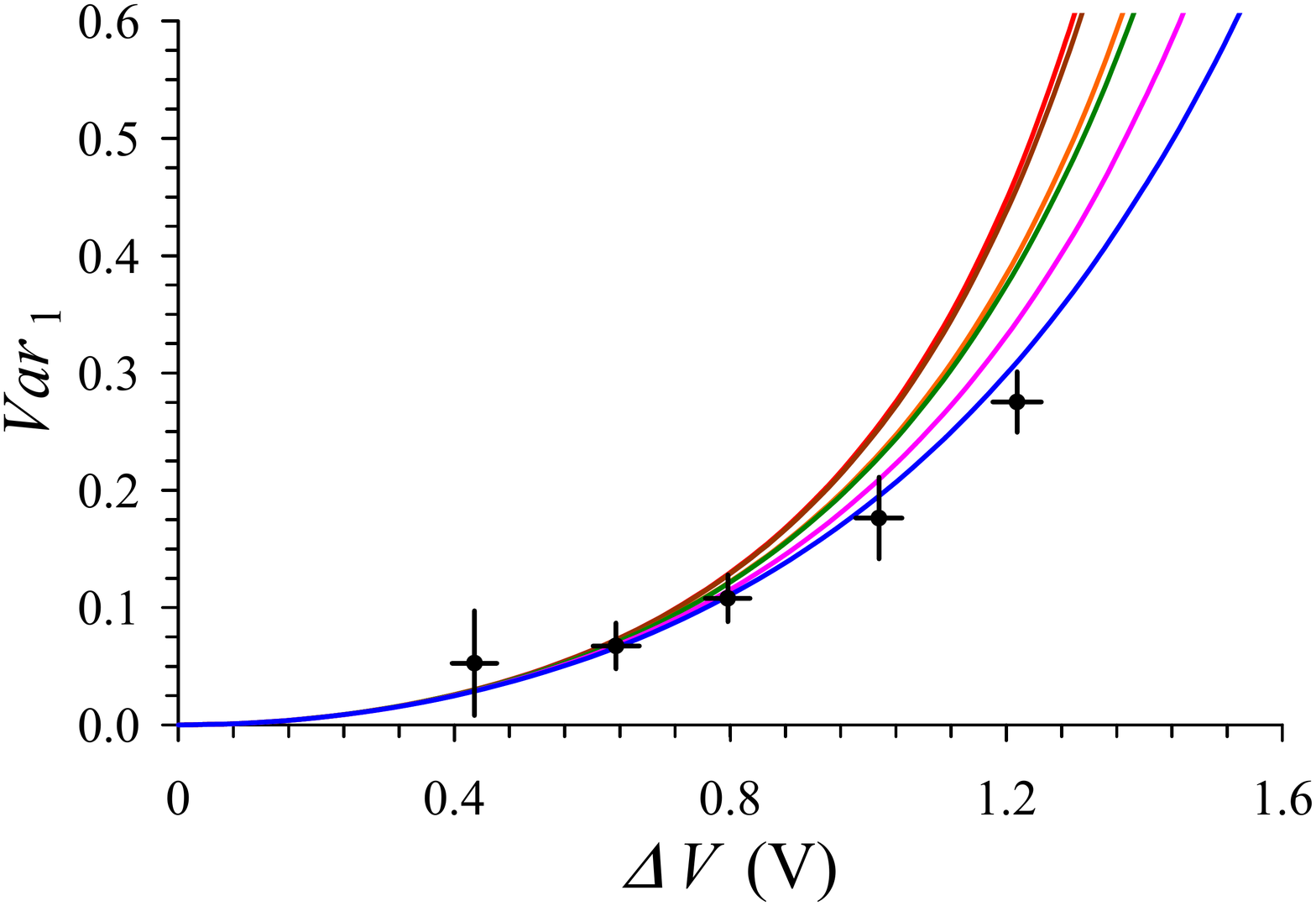}
\end{center}
\caption{Comparison of experimental and theoretical results for the $E_1$ and $Var_1$ parameters.
Theoretical curves are derived using the dependence of $\varepsilon$ on overvoltage given by the fit
of equation (\protect\ref{epsilon_overvoltage}) to data.}
\label{fig:comparison_pulsed_illumination}
\end{figure}

Crosstalk basically contributes to shift and broaden the pulse-height spectrum as described in
section~\ref{ssec:random_primaries}. Experimental $E_1$ and $Var_1$ values were obtained using (\ref{mean_total}) and
(\ref{variance_total}), where the mean $E_{\rm tot}$ and variance $Var_{\rm tot}$ were calculated for the observed
probability distribution, and $E_{\rm pr}=Var_{\rm pr}=\mu$ is already known. The resulting $E_1$ and $Var_1$
parameters are represented in figure~\ref{fig:comparison_pulsed_illumination} by full circles.

Theoretical predictions for the crosstalk models are compared with these experimental $E_1$ and $Var_1$ values. Results
for both the geometric and Borel distributions were calculated using the available analytical expressions
\cite{Vinogradov2}, while for our proposed models they were numerically computed from the probability distributions of
table~\ref{tab:probabilities} extended to $k>5$ using the geometric extrapolation of equation
(\ref{geometric_extrapolation}). The curves of $E_1$ and $Var_1$ versus overvoltage shown in
figure~\ref{fig:comparison_pulsed_illumination} are derived by using the dependence of $\varepsilon$ on overvoltage
obtained from the fit of equation (\ref{epsilon_overvoltage}) to data.

All the crosstalk models yield very similar $E_1$ and $Var_1$ values at $\Delta V<1$~V in agreement with experimental
results. Note that these parameters are model independent up to first order in $\varepsilon$
(see~(\ref{mean_geometrical_models}) and (\ref{variance_geometrical_models})). However,  differences at higher order
become important at $\varepsilon>0.1$. In fact, predictions from models with $n>4$ deviate significantly from
experimental data at $\Delta V>1$~V.

Unlike the measurements at dark-count conditions of section~\ref{ssec:dark_counts}, these results do not allow us to
clearly discriminate between crosstalk models, because the statistics is lower and the systematic errors are more
difficult to control (e.g., photon arrival is assumed to be Poisson distributed with constant mean during the run).
Nevertheless, our experimental data under pulsed-illumination are also compatible with the 4-nearest-neighbors model.

\section{Conclusions}
\label{sec:conclusions}

The effects of optical crosstalk on the photon-counting properties of SiPMs were studied in detail. Novel analytical
models describing the statistics of crosstalk events were developed and compared with experimental results.

The proposed models take into account that each pixel of the array have a finite number of available neighboring pixels
to excite via crosstalk. Effects of cascades due to crosstalk propagation from neighbor to neighbor were included in a
similar way to the models previously reported in \cite{Vinogradov1,Vinogradov2}, which can be regarded as limit
situations of ours. In addition, geometrical considerations on the development of these cascades on the array, which
can lead to saturation of pixels, were accounted for. As a result, analytical expressions for the probability
distribution of the total number of triggered pixels have been reported assuming four different geometrical
arrangements of neighbors: 4 nearest neighbors, 8 nearest neighbors, 8 L-connected neighbors, and all neighbors.

Application of these analytical models to evaluate the effects of crosstalk on the photon statistics under
pulsed-illumination conditions was also discussed. Simple expressions were obtained to account for the shift and
broadening of the spectrum as well as for the excess noise factor due to crosstalk.

A dedicated experiment, using the SiPM device S10362-11-100C from Hamamatsu, was carried out to validate our
theoretical results. Measurements were performed at dark-count conditions as well as under pulsed illumination. A
waveform analysis was developed to obtain the pulse-height spectrum with very high resolution. In particular, the
influence of short-delay afterpulses had to be carefully taken into account. As a by-product of this analysis, the
recovery time of pixels was estimated to be 40~ns at room temperature.

Results of dark-count measurements were found to be consistent with the 4-nearest-neighbor model, in agreement with
other authors that performed MC simulations of the response of SiPMs with crosstalk effects
\cite{SanchezMajos,Vacheret,Dovrat}. The other crosstalk models, including those of \cite{Vinogradov1,Vinogradov2},
failed to match the experimental probability distributions, which demonstrates that the above-mentioned geometrical
considerations are important. Our results indicate that optical crosstalk takes place between adjacent pixels, although
excitation of distant pixels could also be possible in a different chip configuration, as observed in \cite{Rech,Otte}.

Pulse-height spectra were measured under low-intensity pulsed illumination as a function of the SiPM bias voltage. As
expected, both the SiPM gain and sensitivity were found to be nearly proportional to overvoltage, and the crosstalk
probability showed a quadratic growth. Comparison of experimental and theoretical results for pulsed illumination was
done by evaluating the effect of crosstalk on the mean and variance of the spectrum. While all the considered crosstalk
models can only account for experimental results at low overvoltage, the 4-nearest-neighbor model was found to be
consistent with experimental data at any overvoltage, which gives a further support to results obtained at dark-count
conditions.

\section*{Acknowledgements}

This work has been supported by the Spanish Ministerio de Ciencia e Innovación (FPA2009-07772 and CONSOLIDER CPAN
CSD2007-42) and Universidad Complutense de Madrid (UCM 2011: GR35/10-A-910600).

\end{document}